\documentclass[twocolumn,longauth]{aa}
\usepackage{graphicx}
\usepackage{txfonts}
\usepackage{tabularx}
\usepackage{xcolor}
\usepackage{float}
\usepackage{amsmath}
\usepackage{enumitem} % begin{itemize}[leftmargin=-3mm]
\usepackage{chngcntr} % for Figure numbering inside Appendix -- then add \counterwithin{figure}{section}\counterwithin{table}{section} after \appendix
\usepackage{bm}
\usepackage{multirow}
\usepackage{cancel}
\usepackage{verbatimbox}
\usepackage{booktabs} % cmidrule
\usepackage{calc}
\usepackage{array}
\usepackage[iso,german]{isodate}

\usepackage{natbib,twoopt}
\usepackage[breaklinks=true,colorlinks=true,citecolor=blue,urlcolor=blue]{hyperref} %% to avoid \citeads line fills
\bibpunct{(}{)}{;}{a}{}{,}             %% natbib format for A&A and ApJ
\makeatletter
  \newcommandtwoopt{\citeads}[3][][]{\href{http://adsabs.harvard.edu/abs/#3}%
    {\def\hyper@linkstart##1##2{}%
     \let\hyper@linkend\@empty\citealp[#1][#2]{#3}}}
  \newcommandtwoopt{\citepads}[3][][]{\href{http://adsabs.harvard.edu/abs/#3}%
    {\def\hyper@linkstart##1##2{}%
     \let\hyper@linkend\@empty\citep[#1][#2]{#3}}}
  \newcommandtwoopt{\citetads}[3][][]{\href{http://adsabs.harvard.edu/abs/#3}%
    {\def\hyper@linkstart##1##2{}%
     \let\hyper@linkend\@empty\citet[#1][#2]{#3}}}
  \newcommandtwoopt{\citeyearads}[3][][]%
    {\href{http://adsabs.harvard.edu/abs/#3}
    {\def\hyper@linkstart##1##2{}%
     \let\hyper@linkend\@empty\citeyear[#1][#2]{#3}}}
\makeatother

\newlength{\mymincolwidth}

\newcommand{\rom}[1]{{{\uppercase\expandafter{\romannumeral #1}}}}
\newcommand{\romup}[1]{{\textup{\uppercase\expandafter{\romannumeral #1}}}}

\usepackage{listings} % contains \lstinline
\newcommand{\incode}[1]{{\raggedright\lstinline|#1|}}
\newcommand{\incodep}[1]{{\raggedright\lstinline|"#1"|}}

\newcommand{\neutralcarbon}{\ifmmode \text{C\text{\sc{i}}} \else {\sc C\,i}\fi}
\newcommand{\ionizedcarbon}{\ifmmode \text{C\text{\sc{ii}}} \else {\sc C\,ii}\fi}
\newcommand{\neutralhydrogen}{\ifmmode \text{H\text{\sc{i}}} \else {\sc H\,i}\fi}
\newcommand{\ionizedhydrogen}{\ifmmode \text{H\text{\sc{ii}}} \else {\sc H\,ii}\fi}
\newcommand{\carbonmonoxide}{\ifmmode \text{\textsc{CO}} \else {\sc CO}\fi}

\def\HII{\ionizedhydrogen}
\def\CI{\neutralcarbon}
\def\CO{\carbonmonoxide}

\def\alphaCO{\relax\ifmmode%
    \alpha_\mathrm{CO}
    \else{$\alpha_\mathrm{CO}$}\fi%
    }
\def\alphaCI{\relax\ifmmode%
    \alpha_\mathrm{C\text{\sc{i}}}
    \else{\alpha_\mathrm{C\text{\sc{i}}}}\fi%
    }
\def\XCO{\relax\ifmmode%
    X_\mathrm{CO}
    \else{$X_\mathrm{CO}$}\fi%
    }
\def\XCI{\relax\ifmmode%
    X_\mathrm{C\text{\sc{i}}}
    \else{$X_\mathrm{C\text{\sc{i}}}$}\fi%
    }
\def\SNR{\relax\ifmmode%
    \mathrm{S/N}
    \else{$\mathrm{S/N}$}\fi%
    }
\def\RCICO{\relax\ifmmode%
    R_{{\mathrm{C\text{\sc{i}}}}/{\mathrm{CO}}}
    \else{$R_{{\mathrm{C\text{\sc{i}}}}/{\mathrm{CO}}}$}\fi%
    }
\def\tauCO21{\relax\ifmmode%
    \tau_{\mathrm{CO21}}
    \else{$\tau_{\mathrm{CO21}}$}\fi%
    }
\def\tauCI10{\relax\ifmmode%
    \tau_{\mathrm{CI10}}
    \else{$\tau_{\mathrm{CI10}}$}\fi%
    }
\def\XCICO{\relax\ifmmode%
    X_{{\mathrm{C\text{\sc{i}}}}/{\mathrm{CO}}}
    \else{$X_{{\mathrm{C\text{\sc{i}}}}/{\mathrm{CO}}}$}\fi%
    }
\def\RNCINCO{\relax\ifmmode%
    N_{\mathrm{C\text{\sc{i}}}}/N_{\mathrm{CO}}
    \else{$N_{\mathrm{C\text{\sc{i}}}}/N_{\mathrm{CO}}$}\fi%
    }
\def\Tkin{\relax\ifmmode%
    T_{\mathrm{kin}}
    \else{$T_{\mathrm{kin}}$}\fi%
}
\def\Tdust{\relax\ifmmode%
    T_{\mathrm{dust}}
    \else{$T_{\mathrm{dust}}$}\fi%
}
\def\CIabundance{\relax\ifmmode%
    \mathrm{[C\text{\sc{i}}/H_2]}
    \else{$\mathrm{[C\text{\sc{i}}/H_2]}$}\fi%
}
\def\CIcolumndensity{\relax\ifmmode%
    N_{\mathrm{C\text{\sc{i}}}}
    \else{$N_{\mathrm{C\text{\sc{i}}}}$}\fi%
}
\def\COabundance{\relax\ifmmode%
    \mathrm{[CO/H_2]}
    \else{$\mathrm{[CO/H_2]}$}\fi%
}
\def\CIabundance{\relax\ifmmode%
    \mathrm{[C\text{\sc{i}}/H_2]}
    \else{$\mathrm{[C\text{\sc{i}}/H_2]}$}\fi%
}
\def\COcolumndensity{\relax\ifmmode%
    N_{\mathrm{CO}}
    \else{$N_{\mathrm{CO}}$}\fi%
}
\def\CObrightness{\relax\ifmmode%
    I_{\mathrm{CO}}
    \else{$I_{\mathrm{CO}}$}\fi%
}
\def\HydrogenSurfaceDensity{\relax\ifmmode%
    \Sigma_{\mathrm{H_2}}
    \else{$\Sigma_{\mathrm{H_2}}$}\fi%
}
\def\Sigmol{\relax\ifmmode%
    \Sigma_{\mathrm{mol}}
    \else{$\Sigma_{\mathrm{mol}}$}\fi%
}
\def\nHtwo{\relax\ifmmode%
    n_{\mathrm{H_2}}
    \else{$n_{\mathrm{H_2}}$}\fi%
}
\def\dv{\relax\ifmmode%
    {\Delta \text{v}}
    \else{${\Delta \text{v}}$}\fi%
}

\def\Kkms{\mathrm{K\,km\,s^{-1}}}
\def\Mspc2{\mathrm{M_{\odot}\,pc^{-2}}}
\def\Mskpc2{\mathrm{M_{\odot}\,kpc^{-2}}}
\def\kms{\mathrm{km\,s^{-1}}}
\def\Msyr{\mathrm{M_{\odot}\,yr^{-1}}}
\def\Lsun{\mathrm{L_{\odot}}}

\def\Umean{\left<U\right>}

\newcommand{\DefineRemark}[2]{%
\expandafter\newcommand\csname rmk-#1\endcsname{#2}%
}
\newcommand{\Remark}[1]{\csname rmk-#1\endcsname}
\DefineRemark{CI3P13P0}{[\mathrm{C}\,{\text{\sc{i}}}]\;(^{3}P_{1}\textnormal{--}^{3}P_{0})}
\DefineRemark{CI}{\mathrm{C}\,{\text{\sc{i}}}}
\DefineRemark{CI10}{[\mathrm{C}\,{\text{\sc{i}}}]\,(1\textnormal{--}0)}
\DefineRemark{CI21}{[\mathrm{C}\,{\text{\sc{i}}}]\,(2\textnormal{--}1)}
\DefineRemark{CO}{\mathrm{CO}}
\DefineRemark{CO10}{\mathrm{CO}\,(1\textnormal{--}0)}
\DefineRemark{CO21}{\mathrm{CO}\,(2\textnormal{--}1)}
\DefineRemark{CO32}{\mathrm{CO}\,(3\textnormal{--}2)}
\DefineRemark{CO43}{\mathrm{CO}\,(4\textnormal{--}3)}
\DefineRemark{COJ10}{\mathrm{CO}\,(J=1 \to 0)}
\DefineRemark{COJ21}{\mathrm{CO}\,(J=2 \to 1)}
\DefineRemark{COJ32}{\mathrm{CO}\,(J=3 \to 2)}
\DefineRemark{COJ43}{\mathrm{CO}\,(J=4 \to 3)}

\DefineRemark{RCI10CO10}{%
{R}_{\mathrm{C{\text{\sc{i}}}{10}}/\mathrm{CO{10}}}%
}
\DefineRemark{RCI10CO21}{%
{R}_{\mathrm{C{\text{\sc{i}}}{10}}/\mathrm{CO{21}}}%
}
\DefineRemark{RCI10CO32}{%
{R}_{\mathrm{C{\text{\sc{i}}}{10}}/\mathrm{CO{32}}}%
}
\DefineRemark{RCI10CO43}{%
{R}_{\mathrm{C{\text{\sc{i}}}{10}}/\mathrm{CO{43}}}%
}

\DefineRemark{LPrimeCO21}{L^{\prime}_{\Remark{CO21}}}
\DefineRemark{LPrimeCI10}{L^{\prime}_{\Remark{CI10}}}
\DefineRemark{ICO10}{I_{\Remark{CO10}}}
\DefineRemark{ICO21}{I_{\Remark{CO21}}}
\DefineRemark{ICO32}{I_{\Remark{CO32}}}
\DefineRemark{ICO43}{I_{\Remark{CO43}}}
\DefineRemark{ICI10}{I_{\Remark{CI10}}}

\DefineRemark{Kkmspc2}{\mathrm{K\,km\,s^{-1}\,pc^{2}}}

\begin{document}

\title{C~\textsc{i} and CO in Nearby Spiral Galaxies - I. Line Ratio and Abundance Variations at $\sim 200$~pc Scales}

\titlerunning{C~\textsc{i} \& CO - I. Line Ratio and Abundance Ratio}
\authorrunning{D. Liu et al.}

\author{%
    Daizhong Liu \inst{\ref{MPE}}
\and
    Eva Schinnerer \inst{\ref{MPIA}}
\and
    Toshiki Saito \inst{\ref{NAOJ}}
\and 
    Erik Rosolowsky \inst{\ref{UAlberta}}
\and
    Adam Leroy \inst{\ref{OSU}}
\and
    Antonio Usero \inst{\ref{Madrid}}
\and
    Karin Sandstrom \inst{\ref{UCSD}}
\and
    Ralf S.\ Klessen \inst{\ref{HeidelbergZfA},\ref{HeidelbergZfW}}
\and
    Simon C.\ O.\ Glover \inst{\ref{HeidelbergZfA}}
\and
    Yiping Ao \inst{\ref{PMO},\ref{USTC}}
\and
    Ivana Be\v{s}li\'c \inst{\ref{BonnAIfA}}
\and
    Frank Bigiel \inst{\ref{BonnAIfA}}
\and
    Yixian Cao \inst{\ref{MPE}}
\and
    J\'{e}r\'{e}my Chastenet \inst{\ref{UCSD},\ref{Gent}}
\and
    M\'{e}lanie Chevance \inst{\ref{HeidelbergARI},\ref{HeidelbergZfA}}
\and
    Daniel A.\ Dale \inst{\ref{Wyoming}}
\and
    Yu Gao \inst{\ref{XiamenUniv}}\thanks{Deceased.}
\and
    Annie Hughes \inst{\ref{Toulouse}}
\and
    Kathryn Kreckel \inst{\ref{HeidelbergARI}}
\and
    J.~M.~Diederik Kruijssen \inst{\ref{HeidelbergARI}}
\and
    Hsi-An Pan \inst{\ref{TamkangUniv}}
\and
    J\'er\^ome Pety \inst{\ref{IRAM},\ref{SorbonneUniv}}
\and
    Dragan Salak \inst{\ref{HokkaidoUniv},\ref{HokkaidoUnivGrad}}
\and
    Francesco Santoro \inst{\ref{MPIA}}
\and
    Andreas Schruba \inst{\ref{MPE}}
\and
    Jiayi Sun \inst{\ref{McMasterUniv},\ref{TorontoUniv}}
\and
    Yu-Hsuan Teng \inst{\ref{UCSD}}
\and
    Thomas Williams \inst{\ref{MPIA}}
}

\institute{%
    Max-Planck-Institut f\"ur extraterrestrische Physik (MPE), Giessenbachstrasse 1, D-85748 Garching, Germany\\
    \email{dzliu@mpe.mpg.de, astro.dzliu@gmail.com}
    \label{MPE}
\and 
    Max-Planck-Institut f\"ur Astronomie, K\"onigstuhl 17, D-69117 Heidelberg, Germany
    \label{MPIA}
\and
National Astronomical Observatory of Japan, 2-21-1 Osawa, Mitaka, Tokyo, 181-8588, Japan
    \label{NAOJ}
\and
Department of Physics, University of Alberta, Edmonton, AB T6G 2E1, Canada
    \label{UAlberta}
\and
Department of Astronomy, The Ohio State University, 140 West 18th Ave, Columbus, OH 43210, USA
    \label{OSU}
\and
    Observatorio Astron\'{o}mico Nacional (IGN), C/Alfonso XII, 3, E-28014 Madrid, Spain
    \label{Madrid}
\and
    Center for Astrophysics and Space Sciences, Department of Physics, University of California, San Diego, 9500 Gilman Drive, La Jolla, CA 92093, USA
    \label{UCSD}
\and
    Universit\"at Heidelberg, Zentrum f\"ur Astronomie, Institut f\"ur Theoretische Astrophysik, Albert-Ueberle-Str 2, D-69120 Heidelberg, Germany
    \label{HeidelbergZfA}
\and
    Universit\"at Heidelberg, Interdisziplin\"ares Zentrum f\"ur Wissenschaftliches Rechnen, Im Neuenheimer Feld 205, D-69120 Heidelberg, Germany
    \label{HeidelbergZfW}
\and
     Purple Mountain Observatory and Key Laboratory for Radio Astronomy, Chinese Academy of Sciences, Nanjing, China
     \label{PMO}
\and
    School of Astronomy and Space Science, University of Science and Technology of China, Hefei, China
    \label{USTC}
\and
    Argelander-Institut f\"{u}r Astronomie, Universit\"{a}t Bonn, Auf dem H\"{u}gel 71, 53121 Bonn, Germany
    \label{BonnAIfA}
\and
    Sterrenkundig Observatorium, Ghent University, Krijgslaan 281-S9, 9000 Gent, Belgium
    \label{Gent}
\and
    Astronomisches Rechen-Institut, Zentrum f\"{u}r Astronomie der Universit\"{a}t Heidelberg, M\"{o}nchhofstra\ss e 12-14, 69120 Heidelberg, Germany
    \label{HeidelbergARI}
\and
    Department of Physics \& Astronomy, University of Wyoming, Laramie, WY 82071, USA
    \label{Wyoming}
\and
    Department of Astronomy, Xiamen University, Xiamen, Fujian 361005, China
    \label{XiamenUniv}
\and
    CNRS, IRAP, 9 Av. du Colonel Roche, BP 44346, F-31028 Toulouse cedex 4, France; Universit\'{e} de Toulouse, UPS-OMP, IRAP, F-31028 Toulouse cedex 4, France
    \label{Toulouse}
\and
    Department of Physics, Tamkang University, No.151, Yingzhuan Road, Tamsui District, New Taipei City 251301, Taiwan
    \label{TamkangUniv}
\and
    Institut de Radioastronomie Millim\'{e}trique (IRAM), 300 rue de la Piscine, F-38406 Saint-Martin-d'H\`{e}res, France
    \label{IRAM}
\and
    Sorbonne Universit\'{e}, Observatoire de Paris, Universit\'{e} PSL, CNRS, LERMA, F-75014, Paris, France
    \label{SorbonneUniv}
\and
    Institute for the Advancement of Higher Education, Hokkaido University, Kita 17 Nishi 8, Kita-ku, Sapporo, Hokkaido 060-0817, Japan
    \label{HokkaidoUniv}
\and
    Department of Cosmosciences, Graduate School of Science, Hokkaido University, Kita 10 Nishi 8, Kita-ku, Sapporo, Hokkaido 060-0817, Japan
    \label{HokkaidoUnivGrad}
\and
    Department of Physics and Astronomy, McMaster University, 1280 Main Street West, Hamilton, ON L8S 4M1, Canada
    \label{McMasterUniv}
\and
    Canadian Institute for Theoretical Astrophysics (CITA), University of Toronto, 60 St George Street, Toronto, ON M5S 3H8, Canada
    \label{TorontoUniv}
}

\date{Accepted: December 13, 2022}

\abstract{%
We present new neutral atomic carbon [C~\textsc{i}]~(${}^3P_1~\to~{}^3P_0$) mapping observations within the inner $\sim 7$~kpc and $\sim 4$~kpc of the disks of NGC~3627 and NGC~4321 at a spatial resolution of 190~pc and 270~pc, respectively, using the Atacama Large Millimeter/Submillimeter Array (ALMA) Atacama Compact Array (ACA). 
We combine these with the CO(2--1) data from PHANGS-ALMA, and literature [C~\textsc{i}] and CO data for two other starburst and/or active galactic nucleus (AGN) galaxies (NGC~1808, NGC~7469), to study: \textit{a)} the spatial distributions of C~\textsc{i} and CO emission; \textit{b)} the observed line ratio $\RCICO = I_{\mathrm{[C{\normalfont\textsc{i}}]}(1\textnormal{--}0)} / I_{\mathrm{CO}(2\textnormal{--}1)}$ as a function of various galactic properties; and \textit{c)} the abundance ratio of [C~\textsc{i}/CO]. 
We find excellent spatial correspondence between C~\textsc{i} and CO emission and nearly uniform $\RCICO \sim 0.1$ across the majority of the star-forming disks of NGC~3627 and NGC~4321. 
However, $\RCICO$ strongly varies from $\sim 0.05$ at the centre of NGC~4321 to $> 0.2 \text{--} 0.5$ in NGC~1808's starbursting centre and NGC~7469's centre with an X-ray-luminous active galactic nucleus (AGN). 
Meanwhile, $\RCICO{}$ does not obviously vary with $\Umean$, similar to the prediction of photodissociation dominated region (PDR) models. 
We also find a mildly decreasing $\RCICO{}$ with an increasing metallicity over $0.7\text{--}0.85\,\mathrm{Z_{\odot}}$, consistent with the literature. 
Assuming various typical ISM conditions representing giant molecular clouds, active star-forming regions and strong starbursting environments, we calculate the local-thermodynamic-equilibrium radiative transfer and estimate the [C~\textsc{i}/CO] abundance ratio to be $\sim 0.1$ across the disks of NGC~3627 and NGC~4321, similar to previous large-scale findings in Galactic studies. 
However, this abundance ratio likely has a substantial increase to $\sim 1$ and $\gtrsim 1$--5 in NGC~1808's starburst and NGC~7469's strong AGN environments, respectively, in line with the expectations for cosmic-ray dominated region (CRDR) and X-ray dominated region (XDR) chemistry. 
Finally, we do not find a robust evidence for a generally CO-dark, C~\textsc{i}-bright gas in the disk areas we probed. 
}

\keywords{galaxies: ISM --- ISM: molecules --- ISM: atoms --- ISM: abundances --- galaxies: spiral --- galaxies: individual: NGC 3627, NGC 4321, NGC 1808, NGC 7469}

\maketitle

\section{Introduction}
\label{subsec: intro RCICO}

\subsection{Neutral atomic carbon in the interstellar medium}

Neutral atomic carbon (C$^0$ or \neutralcarbon) is an important phase of carbon in the interstellar medium (ISM) besides ionized carbon (C$^+$ or \ionizedcarbon) and carbon monoxide (CO). 
Its $^{3}P$ state is split into three fine-structure levels. The ${}^{3}P_{1} \to {}^{3}P_{0}$ transition line at 492.16065~GHz, 
hereafter $\Remark{CI10}$%
, has an upper energy level of $E_u = 23.620$~K and a critical density of $n_{\mathrm{crit}} \sim 1.0 \times 10^{3} \;\mathrm{cm^{-3}}$ (at a temperature of 50~K)\,\footnote{$n_{\mathrm{crit}} \equiv A / \gamma$, where A is the Einstein A coefficient for the transition and $\gamma$ the rate of collisional de-excitation with hydrogen molecules, with values from \cite{Schroder1991} as compiled in the Leiden Atomic and Molecular Database (\citealt{Schoier2005}), \url{https://home.strw.leidenuniv.nl/~moldata/}}, 
and the ${}^{3}P_{2} \to {}^{3}P_{1}$ transition line at 809.34197~GHz has $E_u = 62.462$~K and $n_{\mathrm{crit}} \sim 3.4 \times 10^{3} \;\mathrm{cm^{-3}}$. 
Therefore, they are easily excited in the cold ISM environment. 
Given the high abundance of carbon in the ISM, \neutralcarbon, CO and \ionizedcarbon{} are the most powerful cold ISM tracers, and are the most widely used tools to probe cold ISM properties and hence galaxy evolution at cosmological distances. 

The interstellar \neutralcarbon{} is produced from
the photodissociation of CO molecules by UV photons and the recombination of \ionizedcarbon{}, both happening in photodissociation regions (PDRs; \citealt{Langer1976,deJong1980,Tielens1985a,Tielens1985b,vanDishoeck1986,vanDishoeck1988,Sternberg1989,Sternberg1995,Hollenbach1991,Hollenbach1999,Kaufman1999,Wolfire2010,Madden2020,Bisbas2021}, also see reviews by \citealt{Genzel1989,Jaffe1985,Hollenbach1997,Wolfire2022}). 
In the simple plane-parallel (i.e., 1-D) PDR model, \neutralcarbon{} exists in a layer where UV photons can penetrate through the molecular gas but cannot maintain a high carbon ionizing rate. 
Such a layer is suggested to have an intermediate gas surface density ($N_{\mathrm{H_2}} \sim 1 \textnormal{--} 4 \times 10^{20} \;\mathrm{cm^{-2}}$), fairly cold temperature (10 to a few tens K), and moderate visual extinction ($A_V \sim 1 - 4$). 
CO molecules dominate the more shielded interior of this layer and \ionizedcarbon{} dominates the exterior. 
Hydrogen molecules are gradually photodissociated to atoms across this layer ($A_V \sim 0.1 - 4$). 
This plane-parallel scenario qualitatively explains the layered structure of the Orion Bar PDR in our Galaxy on sub-pc scales (e.g.,  \citealt{Genzel1989,Tielens1993,Tauber1994,Hogerheijde1995b,Hollenbach1999,Goicoechea2016}), and the $\rho$~Ophiuchi~A PDR in a recent study with an unprecedented resolution of 360~AU (\citealt{Yamagishi2021}), where the interplay between individual young, massive (O, B) stars and the ISM can be directly observed. 

However, such simple models cannot fully explain molecular clouds. Plenty of observations in Galactic molecular clouds, \ionizedhydrogen{} regions and star-formation complexes have revealed a widespread \neutralcarbon{} distribution over most areas of molecular clouds on scales of a few to ten pc, and a relatively high \neutralcarbon{} fractional abundance even at a large $A_V$ depth $> 10$~mag (\citealt{Phillips1980,Phillips1981,Wootten1982,Keene1985,Keene1987,Jaffe1985,Zmuidzinas1986,Zmuidzinas1988,Genzel1988,Frerking1989,Plume1994,Plume1999,Gerin1998a,Tatematsu1999,Ikeda1999,Ikeda2002,Yamamoto2001,Zhang2001,Kamegai2003,Oka2004,Izumi2021}). 

This observational evidence actually requires PDRs to be clumpy  (\citealt{Stutzki1988,Genzel1988,Burton1990,Meixner1993,Spaans1997,Kramer2004,Kramer2008,Pineda2008,Sun2008}), with a volume filling factor much less than unity (e.g., $\sim 0.1 - 0.3$; \citealt{Stutzki1988}), so that 
UV photons can penetrate through most of the cloud, except for the densest clumps. 
Therefore, the spatial co-existence of \neutralcarbon{} and low rotational transition (low-$J$) CO lines can be well explained by such an inter-clump medium inside the molecular clouds. 
Alternatively, cosmic rays (CRs) have also been proposed to be the reason for the \neutralcarbon{}-CO association as they can penetrate much deeper than UV photons into clouds, and thus more uniformly dissociate CO molecules into \neutralcarbon{} inside clouds (\citealt{Field1969,Padovani2009,Papadopoulos2010c,Ivlev2015,Papadopoulos2018}).

\subsection{Variations of the \CI/\CO\ line ratio (\RCICO) in the literature}

Although the spatial distributions of \CI{} and low-$J$ CO emission resemble each other surprisingly well from molecular clouds to Galactic scales (e.g., across the Galactic plane; \citealt{Wright1991,Bennett1994,Oka2005,Burton2015}) and in external galaxies with spatially resolved observations (e.g., \citealt{Schilke1993,White1994b,Harrison1995,Israel1995,Israel2001,Israel2003,Zhang2014,Krips2016,Cicone2018,Crocker2019,Jiao2019,Miyamoto2018,Miyamoto2021,Salak2019,Izumi2020,Saito2020,Michiyama2020,Michiyama2021}), 
their line intensity ratio, $\Remark{RCI10CO10} = \Remark{ICI10} / \Remark{ICO10}$\,\footnote{The line intensity $I$ has brightness temperature units $\mathrm{K\,km\,s^{-1}}$.}, does vary according to local ISM conditions --- \CI{} and CO fractional abundance (i.e., metallicity), excitation condition (e.g., excitation temperature $T_{\mathrm{ex}}$ and optical depth $\tau$), and UV and/or CR radiation field. 
At individual molecular cloud scales, $\Remark{RCI10CO10}$ (or column density ratio) seems to vary from $\gtrsim 0.5$ (or $\gtrsim 0.2$ for column density ratio) to $<0.1$ with an increasing $N_{\mathrm{H2}}$ and $A_V$, e.g., from cloud surface or PDR front to the interior \citep[e.g.][]{Frerking1982,Genzel1988,Oka2004,Kramer2004,Kramer2008,Mookerjea2006,Sun2008,Yamagishi2021}. 

At sub-galactic scales, when individual clouds are smoothed out, $\Remark{RCI10CO10}$ is found to lie in a narrower range of $\sim 0.1 \text{--} 0.3$. For example, across the Galactic plane, the ratio is found to change only slightly within a galactocentric radius of 3--7~kpc ($\sim 0.08 \text{--} 0.12$, overall mean $= 0.105 \pm 0.004$; \citealt{Oka2005}); 
whereas in the starburst galaxies like M~82 and NGC~253, $\Remark{RCI10CO10}$ or the [\CI]/CO abundance ratio is a factor of 3--5 higher than the Galactic value (\citealt{Schilke1993,White1994b,Krips2016}). 
Recent high-resolution (hundred pc scales) $\Remark{CI10}$ mapping studies in the nearby spiral galaxy M~83 (\citealt{Miyamoto2021}) and starburst galaxy IRAS~F18293-3413 (\citealt{Saito2020}) reported tightly-correlated $\Remark{CI10}$ and $\Remark{CO10}$ emission, with scaling relations at hundred parsecs equivalent to
$\Remark{RCI10CO10} = 0.14 \times (L^{\prime}_{\Remark{CO10}}/10^{6.0}\,[\Remark{Kkmspc2}])^{-0.13}$ 
and 
$\Remark{RCI10CO10} = 0.21 \times (L^{\prime}_{\Remark{CO10}}/10^{6.8}\,[\Remark{Kkmspc2}])^{+0.54}$, 
respectively. 
The discrepant power-law indices reflects different trends in individual galaxies that are not fully understood.

Meanwhile, in the central $\lesssim 100$~pc of NGC~7469 close to the X-ray-luminous active galactic nucleus (AGN), \RCICO{} is found to be substantially enhanced by a factor of $>10$ (reaching $\Remark{RCI10CO10} \approx 1$; \citealt{Izumi2020}; see also the Circinus Galaxy's AGN with $\Remark{RCI10CO32} \approx 0.9$ in \citealt{Izumi2018}). 

Furthermore, in the low-metallicity Large Magellanic Cloud (LMC) and Small Magellanic Cloud (SMC) environments, a factor of $1.5 - 3 \times$ higher \RCICO{} values have been found (\citealt{Bolatto2000b,Bolatto2000c}; see also other low-metallicity galaxies: \citealt{Bolatto2000a,Hunt2017}). 

These studies indicate that \RCICO{} is sensitive to ISM conditions among and within galaxies. 
However, previous extragalactic studies are mainly focused on starburst galaxies and bright galaxy centres. How \RCICO{} behaves across the disks and spiral arms of typical star-forming galaxies is much less explored. Whether \CI{} can trace the so-called CO-dark molecular gas at a low gas density or metallicity is also still in question. 
Understanding \RCICO{} variation and quantifying the \CI{} and \CO{} excitation in extragalactic environments is now particularly important given the rapidly growing numbers and diverse galactic environments of \CI{} line observations at cosmological distances (e.g., \citealt{Weiss2003,Weiss2005,Pety2004,Wagg2006,Walter2011,Danielson2011,AlaghbandZadeh2013,Gullberg2016,Bothwell2017,Popping2017,Yang2017,Emonts2018,Valentino2018,Valentino2020,Bourne2019,Nesvadba2019,Dannerbauer2019,Jin2019,Brisbin2019,Boogaard2020,Cortzen2020,Harrington2021,Dunne2021,Lee2021}).

\subsection{What drives \RCICO? --- the need for high-resolution mapping across wide galactic environments}

High-resolution mapping of \CI{} and CO from disks and spiral arms of nearby galaxies is indispensable for understanding and quantifying the ISM \CI{} and CO physics. In particular, the following questions will be answered only with such observations: 
Do \CI{} and low-$J$ CO have the same spatial distributions across galaxy disks? 
Does the \RCICO{} line ratio vary significantly across galaxy disks?
How does the \RCICO{} line ratio vary among different local galaxies?
How can the observed \RCICO{} line ratio be reliably converted to the \CI{} and CO column density ratio (i.e., solving their radiation transfer equations and level population statistics and obtaining optical depths and excitation temperature)?

Currently, high-resolution (hundreds of parsec scale) imaging/mapping of \CI{} in nearby galaxies relies on radio telescopes at high altitudes, e.g., the Atacama Large Millimeter/submillimeter Array (ALMA), Atacama Submillimeter Telescope Experiment (ASTE, 10-m single-dish), Atacama Pathfinder Experiment (APEX, 12-m single-dish), and is limited to the available weather conditions corresponding to the high frequency of \CI{} lines. Therefore, only a handful of galaxies have been imaged/mapped. 
These include NGC~6240 by \citet{Cicone2018}, Circinus Galaxy by \citet{Izumi2018}, NGC~613 by \citet{Miyamoto2018}, NGC~1808 by \citet{Salak2019}, NGC~7469 by \citet{Izumi2020}, IRAS~F18293-3413 by \citet{Saito2020}, NGC~6052 by \citet[][\CI{} undetected]{Michiyama2020}, 36 (U)LIRGs by \citet{Michiyama2021}, M~83 by \citet[][mapped the centre and northern part with ASTE]{Miyamoto2021}, Arp~220 by \citet{Ueda2022}, and NGC~1068 by \citet{Saito2022a,Saito2022b}. 
Nevertheless, all of these observations, except for M~83, targeted the centres of starburst, IR-luminous and/or AGN host galaxies. M~83 is the only typical star-forming main sequence galaxy having a high-quality \CI{} mapping. However, its map is still limited to a quarter of its inner $\sim$3~kpc disk. 

In this work, we present two new \CI{} mapping observations with the ALMA Atacama Compact Array (or Morita Array; 7-m dish, hereafter ACA or 7m) in nearby spiral galaxies NGC~3627 and NGC~4321.
Our observations aimed at detecting \CI{} in the disks, and thus are much larger-area and deeper than previous \CI{} maps. 
By further adding a starburst and an AGN-host galaxy from the literature into our analysis, we present a comprehensive study of the $\Remark{RCI10CO21}$ line ratio (hereafter $\RCICO$; ratios of other \CI{}/CO transitions will be explicitly written otherwise) in nearby star-forming disk and starburst and AGN environments. 
In order to quantify the \CI{} and CO abundance ratio, 
we have further calculated the local thermodynamic equilibrium (LTE) and non-LTE radiative transfer of \CI{} and CO under several \textit{representative ISM conditions}. 
The results from this study could thus be one of the most comprehensive local benchmarks for understanding the \CI{} and CO line ratio and abundance variations. 

The paper, as the first one of our series of papers, is organized as follows. 
Sect.~\ref{sec: Observations} describes the sample selection, observations and data reduction. 
Sect.~\ref{sec: Results} presents the spatial distribution and variation of $\RCICO$. 
Sect.~\ref{sec: Excitation Analysis} presents the analysis of \CI{} and CO excitation and radiative transfer and thus the abundance (species column density) ratio. 
In Sect.~\ref{sec: Discussion} we discuss various topics relating to the [\CI/CO] abundance variation, CRDR, XDR, and CO-dark gas. 
Finally we conclude in Sect.~\ref{sec: Conclusion}. 
Our companion paper~II (D. Liu et al. in prep.) presents the detailed radiative transfer calculation and \CI{} and CO conversion factors, and is related to the abundance calculation in this paper.

\vspace{2ex}

\section{Targets and observations}
\label{sec: Observations}

\subsection{Target selection}
\label{subsection: Target Selection}

Our targets, NGC~3627 and NGC~4321, are selected from a joint sample of the PHANGS-ALMA CO(2--1) (\citealt{Leroy2021phangs}), PHANGS-HST (\citealt{PHANGSHST}), and PHANGS-MUSE (\citealt{PHANGSMUSE}) surveys, with CO line intensities and systemic velocities best suitable for ALMA Band 8 \CI{} mapping. 

The PHANGS-ALMA sample consists of 90 nearby star-forming galaxies initially selected at distances of $2 \lesssim d / \mathrm{Mpc} \lesssim 23$, with a stellar mass $\log (M_{\star}/\mathrm{M_{\odot}}) > 9.75$, and which are not too inclined and visible to ALMA. They represent the typical star-forming main sequence galaxies in the local Universe. 
The PHANGS-HST sample \citep{PHANGSHST} is a subsample of 38 PHANGS-ALMA galaxies, providing high-resolution stellar properties. 
The PHANGS-MUSE large program \citep{PHANGSMUSE} targeted a subsample of 19 PHANGS-ALMA galaxies suitable for VLT/MUSE optical integral field unit (IFU) observations, achieving similar spatial coverage as in PHANGS-ALMA and providing rich nebular emission lines, attenuation, stellar age and metallicity and H~\textsc{ii} region information (e.g., \citealt{Kreckel2019, Kreckel2020, Pessa2021, Santoro2021, Williams2022}). 

For the legacy value, we selected our sample from the joint ALMA+MUSE+HST sample pool, considered CO surface brightness, mapping area, \CI{} line frequency and the transmission in the ALMA Band 8. 
In this work, we present the new \CI{} data of NGC~3627 and NGC~4321.
Because of the ALMA Band 8 sensitivity and the expected fainter \CI{} line intensity than CO, the \CI{} mapping areas are smaller than their CO(2--1) maps, but they are still the largest ($\sim1'-2.35'$) and deepest (rms~$\sim0.04$~K per $\sim5$~km~s$^{-1}$) \CI{} mapping with ALMA. 

Furthermore, in order to study the \CI{} and CO in different galactic environments, we selected two starbursting and/or AGN-host galaxies from the literature whose ALMA $\Remark{CI10}$ and CO(2--1) data are available in the archive: NGC~1808 \citep{Salak2019} and NGC~7469 \citep{Izumi2020}. 
NGC~1808 has ALMA mosaic observations covering its $\sim 1$~kpc starburst ring, and NGC~7469 has ALMA single-pointing observations covering its inner $\sim 3$~kpc area.

\subsection{New ALMA ACA Band 8 observations}
\label{subsection: ALMA Band 8 Observation}

Our \CI{} observations for NGC~3627 were taken under the ALMA project 2018.1.01290.S (PI: D. Liu) during June 3rd to Sept. 29th, 2019 with ACA-only at Band~8. The total on-source integration time was 30.8~hours. A mosaic with 149 pointings was used to map an area of $94'' \times 136''$ at a position angle of $-27^\circ$, which covers the galaxy centre, bar and the majority of the spiral arms, and is about 1/3 of the full PHANGS-ALMA CO(2-1) map area. 
The achieved line sensitivity per $5\,\kms$ is 45--55~mK (107--130~$\mathrm{mJy/beam}$) across the $\Remark{CI10}$ line frequencies with a beam of 3.46$''$. 

The \CI{} mapping for NGC~4321 were taken under the ALMA project 2019.1.01635.S (PI: D. Liu; NGC~1365 is also partially observed for this project) during Dec. 3rd, 2019 to July 1st, 2021, also with ACA-only at Band~8. The on-source integration time reached 16.1~hours for the mapping of an area of $55'' \times 42''$ with 23 pointings using the 7m array. The achieved line sensitivity is similar to that of NGC~3627. 
More detailed information is provided in Table~\ref{table: targets}. 

ACA (7m) was chosen to provide a matched and slightly coarser synthesized beam ($\sim 3''-4''$) than the PHANGS-ALMA CO(2-1) data. In comparison, the main 12m ALMA array in compact C43-1/2 configurations will result in a $\sim 0.4''-0.7''$ beam and significantly miss large scale ($>4-6''$) emission without ACA. 

No total power observations were requested in our presented observation, therefore we matched the $uv$ ranges of the raw CO(2--1) and $\Remark{CI10}$ visibility data during our data reduction. 
We verified that missing flux will not substantially ($\sim 30\%$) bias our analysis using three methods as detailed in Appendix~\ref{appendix: missing flux}. This includes a CASA simulation of visibilities mimicking our ACA mosaic observations in NGC~3627, from which we find a missing flux as small as 7\% in \CI-bright pixels and up to $\sim 50$\% in the faintest pixels (and mostly $< 30\%$).

Our data reduction follows the PHANGS-ALMA imaging and post-processing pipeline \citep[][]{Leroy2021pipeline}, including imaging and deconvolution, mosaicking, broad and strict/signal mask generation, and moment map creation. 
The additional steps in this study are: (a) $uv$-clipping in the CO(2--1) data to match the $uv$ sampling range of $\Remark{CI10}$; and (b) making a joint mask of the individual CO(2--1) and $\Remark{CI10}$ masks then extracting the moment maps (in the joint broad mask) and line ratios maps (in the joint strict mask). 
More technical details are provided in Appendix~\ref{appendix: data reduction}. 

\begin{figure*}[htb]
\centering%
\includegraphics[width=\textwidth, trim=0 0 0 0]{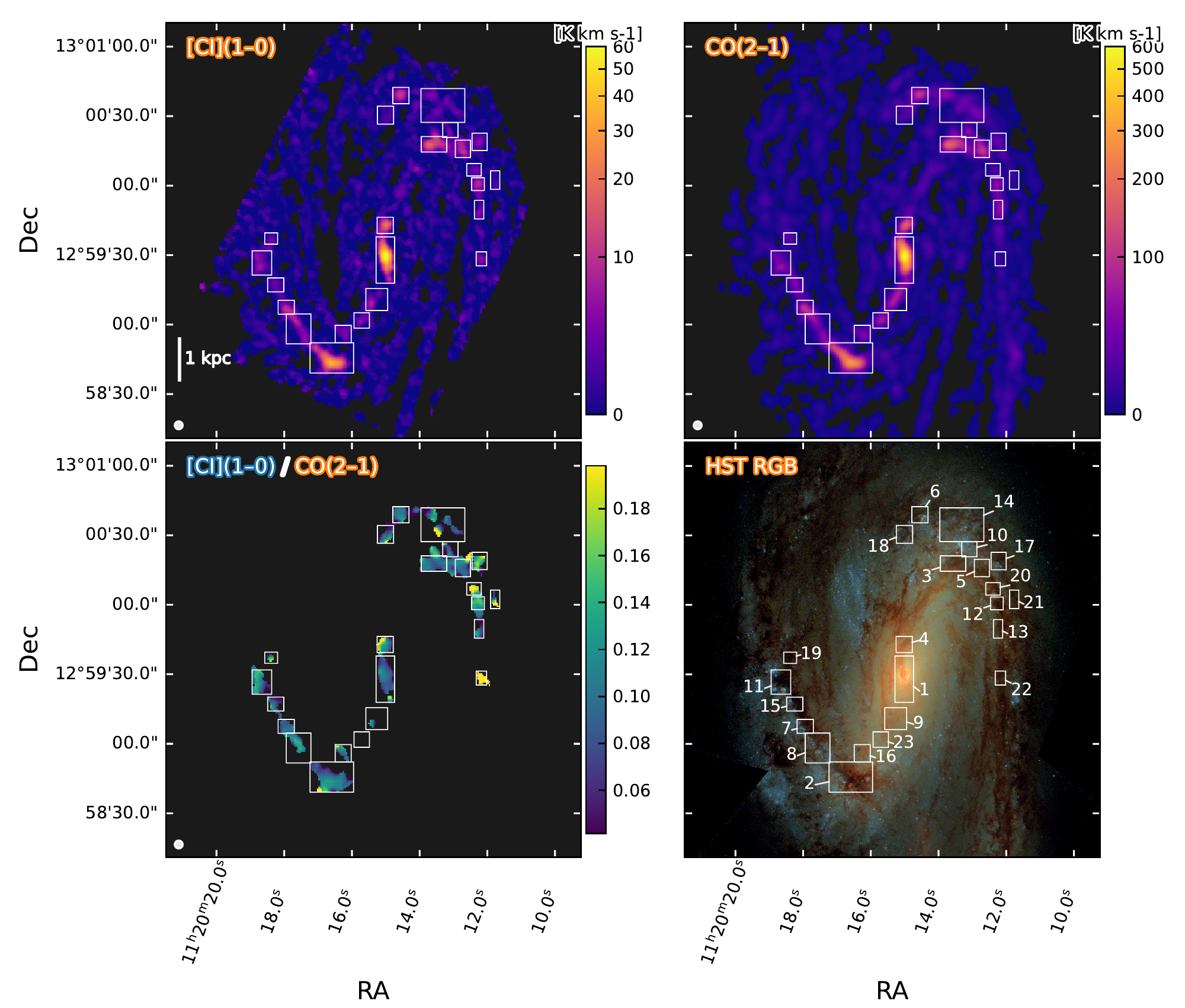}
\caption{\textit{Top panels}: NGC~3627 $\Remark{CI10}$ (\textit{left}) and $\Remark{CO21}$ (\textit{right}) integrated line intensity (moment-0) maps within our broad mask (representing high-completeness signals; see Appendix~\ref{appendix: data reduction}). 
\textit{Bottom left panel}: $\Remark{CI10}/\Remark{CO21}$ line ratio map within the combined signal mask. 
\textit{Bottom right panel}: HST F814W-F555W-F438W RGB composite image (PHANGS-HST; \citealt{PHANGSHST}). 
Regions marked with the white boxes and labeled with numbers are mainly for illustration purpose and for discussion in the text. Their zoom-in images and extracted spectra are provided in our online data release. Examples for Regions~1, 4 and 22 are presented in Appendix~\ref{appendix: all regions}. 
All images have the same field of view of $179'' \times 179''$, or $9.83\,\mathrm{kpc} \times 9.83\,\mathrm{kpc}$. 
}
\label{fig: NGC3627 line ratio map}
\end{figure*}

\begin{figure*}[htb]
\centering%
\includegraphics[width=\textwidth, trim=0 0 0 0]{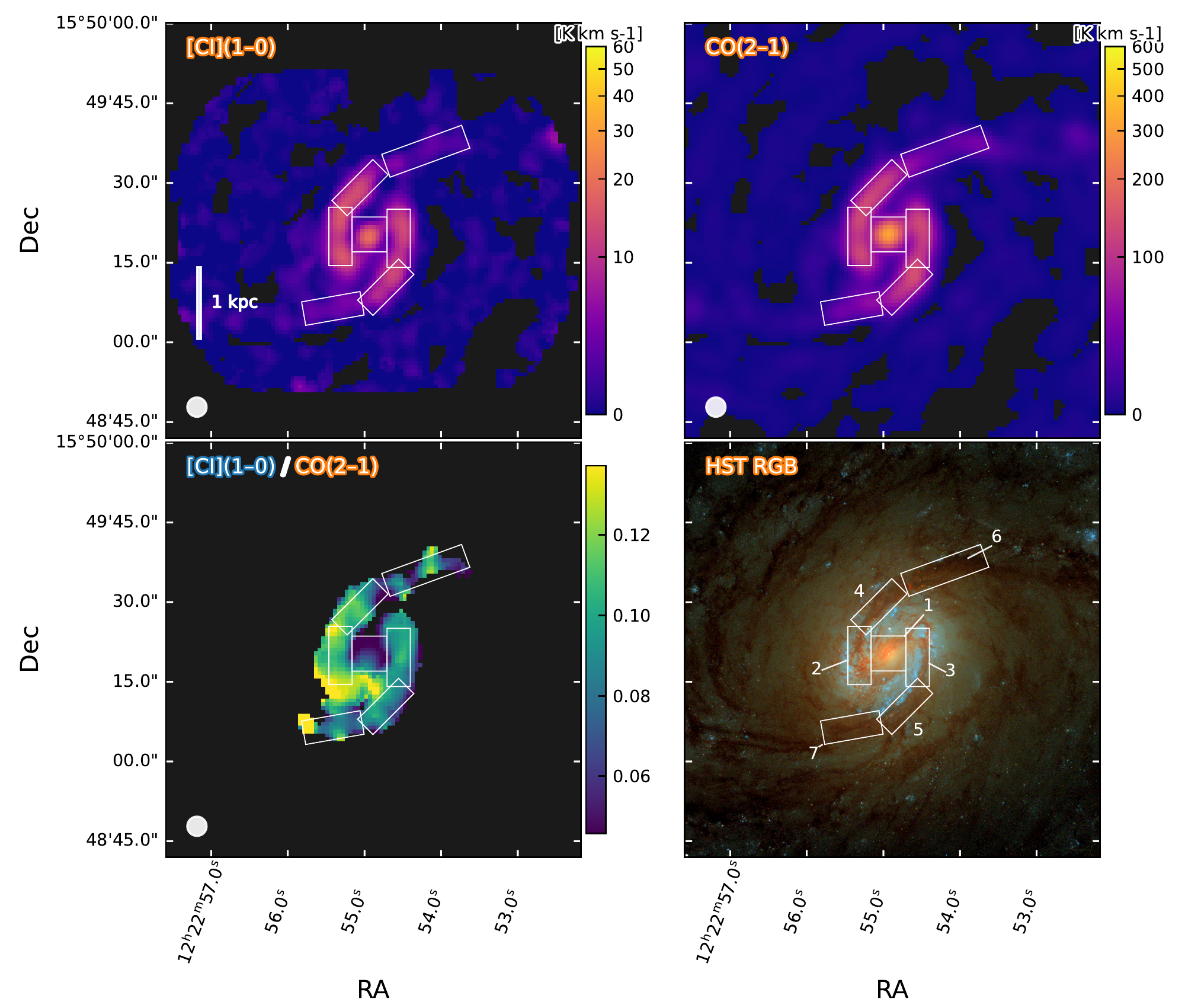}
\caption{%
Similar to Fig.~\ref{fig: NGC3627 line ratio map}, the $\Remark{CI10}$ and CO(2--1) moment-0 intensity maps and the line ratio map and the HST composite RGB image for NGC~4321. The field of view is $78'' \times 78''$, or $5.75\,\mathrm{kpc} \times 5.75\,\mathrm{kpc}$. 
}
\label{fig: NGC4321 line ratio map}
\end{figure*}

Our $\Remark{CI10}$ and CO(2--1) line intensity maps, the line ratio maps and \textit{HST} images for the two PHANGS galaxies, NGC~3627 and NGC~4321, are presented in Figs.~\ref{fig: NGC3627 line ratio map} and \ref{fig: NGC4321 line ratio map}, respectively. 
Our data cubes and moment maps are also made publicly available at 
the PHANGS-ALMA CADC public repository\,\footnote{\url{https://www.canfar.net/storage/list/phangs/RELEASES/DZLIU_etal_2022}}.

\begin{table*}[htb]
    \centering
    \caption{\label{table: targets}%
        Physical and Data Properties of the Sample
    }
    \begin{tabularx}{\linewidth}{l c c c c}
        \setlength{\mymincolwidth}{\dimexpr 0.85\linewidth/5 \relax}
        \global\mymincolwidth=\mymincolwidth
        \rule{\mymincolwidth}{0pt} & 
            \rule{\mymincolwidth}{0pt} & 
            \rule{\mymincolwidth}{0pt} & 
            \rule{\mymincolwidth}{0pt} & 
            \rule{\mymincolwidth}{0pt} 
            \\[-\arraystretch\normalbaselineskip]
        \hline
        \hline
         --- & 
            NGC\,3627 & 
            NGC\,4321 & 
            NGC\,1808 & 
            NGC\,7469 \\
         \hline
         redshift & 
            0.00243 & 
            0.00524 & 
            0.00332 & 
            0.01627 \\
         distance (Mpc) & 
            11.32 & 
            15.21 & 
            10.8 & 
            71.2 \\
         linear scale (pc/$''$) & 
            54.9 & 
            73.7 & 
            52.4 & 
            345.2 \\
         angular resolution & 
            3.46$''$ (190\,pc) & 
            3.65$''$ (269\,pc) & 
            2.67$''$ (140\,pc) & 
            0.46$''$ (159\,pc) \\
         channel width ($\mathrm{km\,s^{-1}}$) & 
            4.77 & 
            4.76 & 
            9.52 & 
            4.76 \\
         \textnormal{[}C\,\textsc{i}\textnormal{]}(1--0) flux scale ($\mathrm{K\,Jy^{-1}}$) & 
            0.42 & 
            0.38 & 
            0.71 & 
            23.8 \\
         CO(2--1) flux scale ($\mathrm{K\,Jy^{-1}}$) & 
            1.92 & 
            3.70 & 
            3.24 & 
            108.4 \\
         \textnormal{[}C\,\textsc{i}\textnormal{]}(1--0) channel rms (K) & 
            0.044 & 
            0.043 & 
            0.056 & 
            0.11 \\
         CO(2--1) channel rms (K) & 
            0.022 & 
            0.015 & 
            0.083 & 
            0.14 \\
         imaging area & 
            $2.35' \times 1.60'$ & % PA 27$^{\circ}$
            $1.28' \times 1.00'$ & 
            $40'' \times 40''$ & 
            $\textnormal{r}\sim9''$ 
            \\
         imaging type & 
            7m, mosaic &
            7m, mosaic & 
            12m+7m+tp, mosaic & 
            12m+7m, single 
            \\
         $M_{\star}$ ($10^{10} \, \mathrm{M_{\odot}}$) & 
            6.8 &
            5.6 &
            2.0 &
            10.7
            \\
         $L_{\mathrm{FIR}}$ ($10^{10} \, \mathrm{L_{\odot}}$) & 
            1.7 & 
            1.9 & 
            3.5 & 
            25 
            \\
         $L_{\mathrm{TIR}}$ ($10^{10} \, \mathrm{L_{\odot}}$) & 
            2.4 & 
            2.5 & 
            5.1 & 
            39 
            \\
         $L_{X}$ ($10^{40} \, \mathrm{erg\,s^{-1}}$) & 
            0.032 & % 0.3\textnormal{--}8\mathrm{keV}, Grier2011
            0.12 & % 0.3\textnormal{--}8\mathrm{keV}, Grier2011
            1.6 & % 2\textnormal{--}10\mathrm{keV}, Jimenez-Bailon 2005
            1500 % 2\textnormal{--}10\mathrm{keV}, Izumi2020, Liu2014
            \\
         SFR ($\mathrm{M_{\odot}\,yr^{-1}}$) & 
            3.84 & 
            3.56 & 
            7.9 (SB ring: 4.5) & 
            76 % Izumi 2020: NGC 7469 SB ring SFR ~ 30--50 (Genzel et al. 1995; Pereira-Santaella et al. 2011)
            \\
         $M_{\mathrm{H_2}}$ ($10^{9} \, \mathrm{M_{\odot}}$) & 
            6.0 & 
            7.8 & 
            (SB ring: 0.52) & 
            15 
            \\
         $M_{\mathrm{H}{\textnormal{\textsc{i}}}}$ ($10^{9} \, \mathrm{M_{\odot}}$) & 
            1.1 & 
            2.7 & 
            1.7 & 
            5.7 
            \\
         mean \RCICO{} & % \tablenotemark{m}
            0.11 & 
            0.10 & 
            0.16 & 
            0.21 
            \\
         mean $\log_{10} \RCICO{}$ & 
            $-0.96$ & 
            $-1.02$ & 
            $-0.83$ & 
            $-0.69$ 
            \\
         $\pm 2 \sigma$ range of $\log_{10} \RCICO{}$ & 
            $-1.26$, $-0.66$ & 
            $-1.24$, $-0.80$ & 
            $-1.03$, $-0.63$ & 
            $-0.95$, $-0.43$ 
            \\
        \hline
        \hline
    \end{tabularx}
    \tablefoot{%
        Redshifts are taken from the NASA Extragalactic Database (NED). 
        The distances of NGC~3627, NGC~4321, NGC~1808 and NGC~7469 are taken from \cite{2009AJ....138..332J}, \cite{2001ApJ...553...47F}, \cite{Salak2019} and \cite{Izumi2020}, %and \cite{Saito2020}, 
        respectively. 
        The stellar masses, SFRs, $M_{\mathrm{H_2}}$ and $M_{\mathrm{H}{\textnormal{\textsc{i}}}}$ of NGC~3627 and NGC~4321 are taken from the PHANGS sample table v1.6 of \citet[][PHANGS survey]{Leroy2021phangs}. 
        The stellar masses and SFRs of NGC~1808 and NGC~7469 are taken from the z0MGS survey (\citealt{Leroy2019}; \url{https://irsa.ipac.caltech.edu/data/WISE/z0MGS/}). 
        The $M_{\mathrm{H_2}}$ and $M_{\mathrm{H}{\textnormal{\textsc{i}}}}$ of NGC~7469 are adopted from \cite{Papadopoulos2000}.
        The global $M_{\mathrm{H}{\textnormal{\textsc{i}}}}$ mass in NGC~1808 is measured by \citet{Koribalski1993}. 
        $L_{\mathrm{FIR}}$ and $L_{\mathrm{TIR}}$ are IR 40--400$\mu$m and 8--1000$\mu$m luminosities all from \citet{Sanders2003}. 
        The X-ray luminosities for NGC~3627 and NGC~4321 are both $L_{0.3\textnormal{--}8\mathrm{keV}}$ from \citet{2011ApJ...731...60G}, and those for NGC~1808 and NGC~7469 are $L_{2\textnormal{--}10\mathrm{keV}}$ from \citet{2005A&A...442..861J} and \cite{Izumi2020}, respectively. 
    }
\end{table*}

\subsection{ALMA observations of the additional targets from the literature}
\label{subsection: additional targets}

The NGC~1808 ALMA $\Remark{CI10}$, $\Remark{CO10}$ and $\Remark{CO21}$ observations are presented in \citet{Salak2019} and all are interferometry plus total power, i.e., 12m+7m+tp, thus having no missing flux issue. 
To study the \CI{}/\CO{} line ratio consistently with our main targets, we only used the $\Remark{CI10}$ and $\Remark{CO21}$ data. 
We re-reduced the raw data under ALMA project 2017.1.00984.S (PI: D. Salak) with the observatory calibration pipeline, then imaged, primary-beam-corrected and short-spacing-corrected using the PHANGS-ALMA pipeline in the same way as for our main targets (Appendix~\ref{appendix: data reduction}). 
Note that in this step we re-reduced and imaged the total power data with the PHANGS-ALMA singledish pipeline (\citealt{Herrera2020,Leroy2021pipeline}) and did the short-spacing correction with the PHANGS-ALMA pipeline right after the primary beam correction. 
The properties of the final \CO{}-\CI{} beam-matched data cubes are listed in Table~\ref{table: targets}. 

The NGC~7469 ALMA data are presented in \citet{Izumi2020} and we used the archival raw data of $\Remark{CI10}$ and $\Remark{CO21}$ under ALMA project 2017.1.00078.S (PI: T. Izumi). These are single pointings toward the centre. $\Remark{CI10}$ is observed with ACA 7m and $\Remark{CO21}$ with the 12m array, both without total power. We re-reduced the data following the same procedure as for NGC~3627 and NGC~4321 with the PHANGS-ALMA pipeline.

\begin{figure*}
\centering%
\includegraphics[width=0.9\textwidth, trim=0 7.7mm 0 2mm, clip]{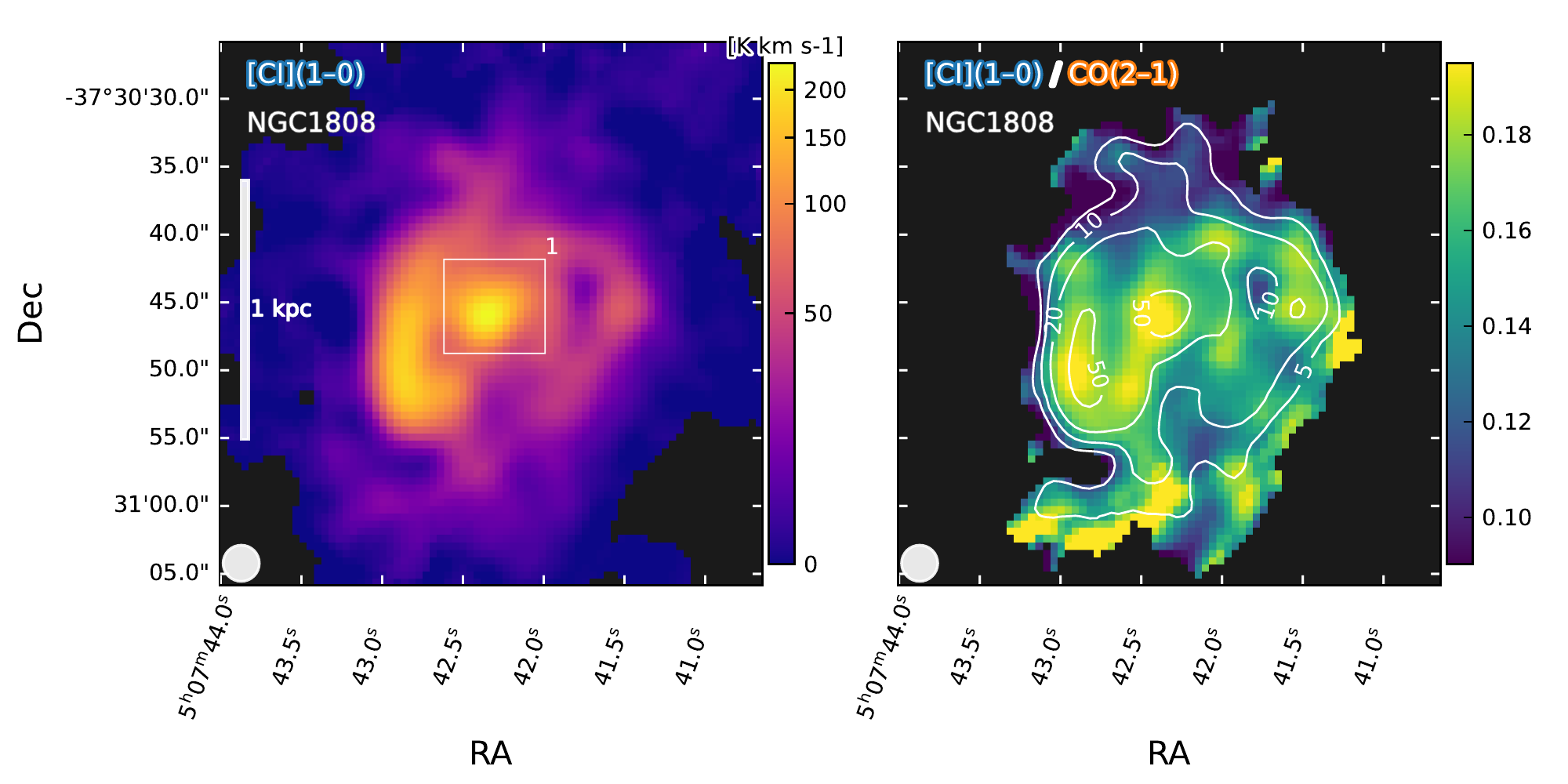}
\includegraphics[width=0.9\textwidth, trim=0 0 0 2mm, clip]{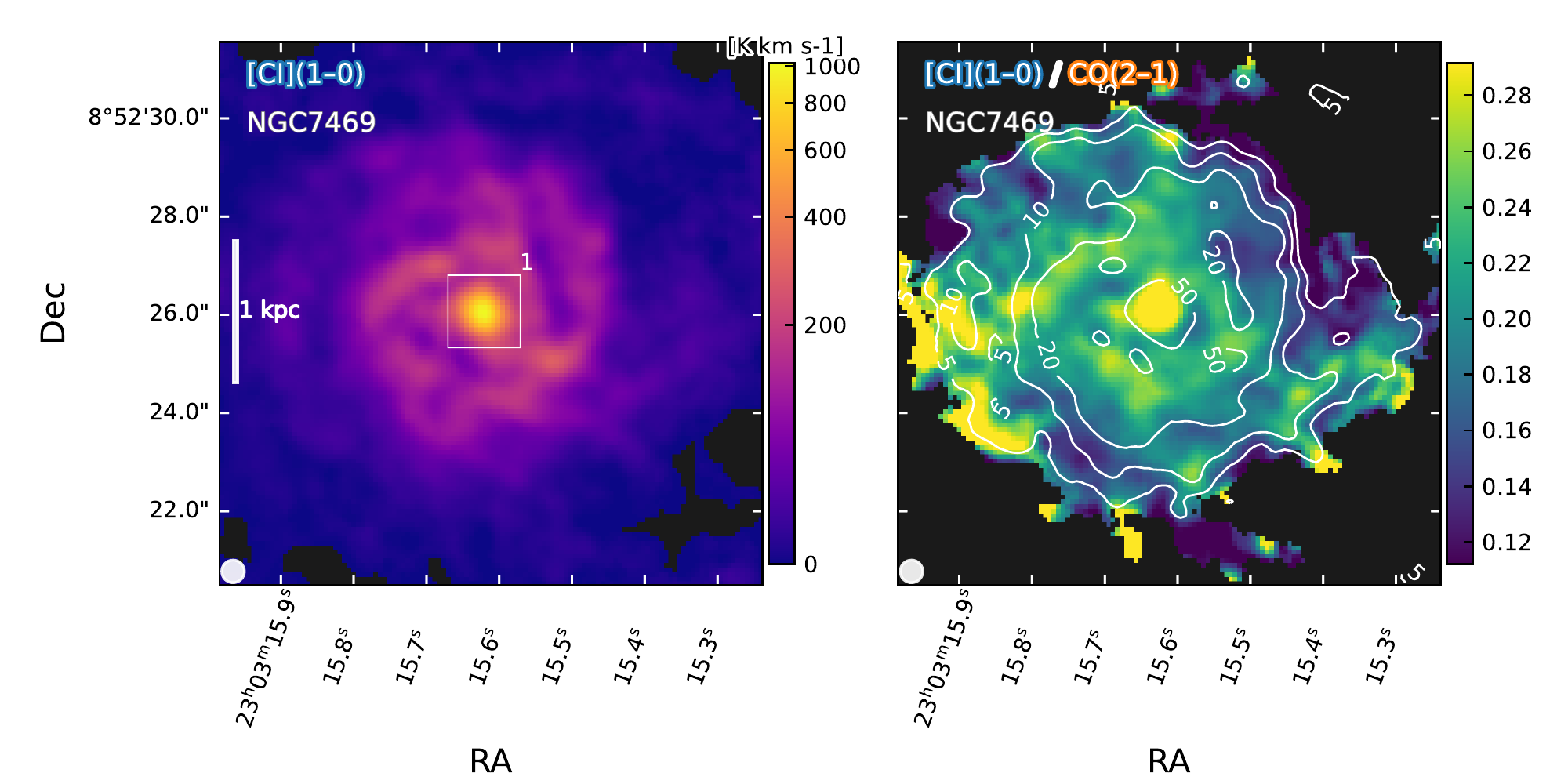}
\caption{%
$\Remark{CI10}$ (\textit{left}) and \RCICO{} (\textit{right}) maps of NGC~1808 in the upper panels and NGC~7469 in the lower panels. 
Similar to Figs.~\ref{fig: NGC3627 line ratio map} and \ref{fig: NGC4321 line ratio map}, the line intensity maps are defined in the broad mask and the \RCICO{} maps are restricted to the combined signal mask. 
Boxes in the left panels indicate the manually selected central regions where the pixels are highlighted in Fig.~\ref{fig: line ratio vs CO}.
Contours in the right panels indicate the 5, 10, 20 and 50-$\sigma$ levels of the left-panel \CI{} intensity maps.
\label{fig: other line ratio maps}
}
\end{figure*}

Our re-reduced $\Remark{CI10}$ and CO(2--1) line intensity and ratio maps for these additional galaxies are shown in Fig.~\ref{fig: other line ratio maps}.

\vspace{2ex}

\section{Results of the observed line ratios}
\label{sec: Results}

\subsection{Spatial distributions}

In Figs.~\ref{fig: NGC3627 line ratio map}, \ref{fig: NGC4321 line ratio map} and \ref{fig: other line ratio maps}, we present the beam-matched $\Remark{CI10}$ and $\Remark{CO21}$ integrated line intensity maps and their ratio maps. 
The \CI/\CO{} line ratio map is computed only for pixels within the joint strict mask area of the two lines, as described in Appendix~\ref{appendix: data reduction}, representing high-confidence signals. 

In NGC~3627 (Fig.~\ref{fig: NGC3627 line ratio map}), we visually marked 23 regions based on the \CI{} and CO commonly-detected ($\mathrm{S/N}>3$), strict-mask areas for later analysis. Region IDs are sorted by their $\Remark{CI10}$ brightness. 
Region 1 corresponds to the galaxy centre, Regions 2 and 3 are the southern and northern bar ends, Region 4 is an offset peak next to the galaxy centre, and other regions are mostly on the spiral arms. Most of these regions show a similar $\RCICO \sim 0.1$, except for, e.g., Regions 20 and 22 which exhibit a somewhat higher ratio of $\RCICO \sim 0.5$. The edge of Region 4 also shows an enhancement of \CI{} yet it is hard to tell whether this is physical or simply caused by edge effects, e.g., as seen in our simulations (Appendix~\ref{appendix: simulation}). 

We also examined the integrated spectra of each region in Appendix~\ref{appendix: all regions}. 
The \CI{} and CO line profiles are not always well-matched in width and shape. In the galaxy centre (Region 1) and at bright bar-ends (Regions 2, 3, 5), \CI{} and CO have similar line widths but \CI{} has a lower intensity than CO, whereas in some fainter regions (9, 10, 13, 14), \CI{} is narrower than CO and relatively weaker. In Regions 4, 20 and 22 where $\RCICO$ is high, \CI{} and CO have similar line widths but the \CI{} intensity is enhanced. 

In NGC~4321 (Fig.~\ref{fig: NGC4321 line ratio map}), we mapped a smaller area due to its fainter CO brightness than NGC~3627. There are no regions with $\Remark{CO21}$ integrated intensities $>100 \,\Kkms$ outside our selected area in NGC~4321, whereas in comparison the NGC~3627 bar ends have $\Remark{CO21}$ integrated intensities $\gtrsim 300 \,\Kkms$. The \CI{} mapping in NGC~4321 has a similar depth as in NGC~3627.

We marked 7 regions in NGC~4321 to guide the visual inspection. A clear \RCICO{} deficit can be seen in the galaxy centre (Region 1), with $\RCICO \sim 0.06$, nearly a factor of two smaller than its inner disk which has similar $\RCICO \sim 0.1$ as seen in the NGC~3627 disk. Some \CI{} enhancement can be found at the edges or slightly outside these regions, where CO and \CI{} are both faint ($\Remark{ICO21} \sim 30-100 \,\Kkms$) and the edge effect of interferometric missing flux might play a role (see Appendix~\ref{appendix: simulation}). 

The \RCICO{} spatial distributions of the other two starburst/AGN galaxies (Fig.~\ref{fig: other line ratio maps}) show in general more enhanced \RCICO{} than in the NGC~3627 and NGC~4321 disks.
Large variations from the centres to outer areas can be seen, especially at the the S/N threshold of $\sim 3 \text{--} 5$ (see contours).

To quantify the spatial co-localization of the \CI{} and CO emission, we use the \incode{JACoP} tool \citep{JACoP}\,\footnote{\url{https://imagej.net/plugins/jacop}}
in the \incode{ImageJ} software \citep{ImageJ} to measure the commonly used co-localization indicators, namely the Pearson's coefficient (PC; equal to 0 and 1 for null and 100\% localization, respectively) and Manders' coefficients (M1 and M2; which measure the overlap between two images and are equal to one for perfect overlap). We turn on the option of using Costes' automatic threshold \citep{Costes2004} which automatically set a pixel value threshold for each image to minimize the contribution of noise for the comparison \citep{JACoP}. 
For the \CI{} and CO emission in our galaxies, we find $\mathrm{PC/M1/M2}$ values all very close to 1, that is, nearly 100\% co-localization ($\mathrm{PC/M1/M2} = 0.934/0.985/0.990$ for NGC~3627, $0.816/0.986/0.964$ for NGC~4321, $0.978/0.890/0.906$ for NGC~1808, and $0.935/0.999/0.998$ for NGC~7469).

Overall, the spatial comparison of the \CI{} and CO intensities demonstrates that the two species are well-correlated when observed at hundred parsec scales. 
Meanwhile, we do see certain spectral differences in the line profiles as shown in Appendix~\ref{appendix: all regions}. This may suggest that variations exist at smaller scales than our resolution, which could be the sign of different evolutionary stages \citep[e.g.][]{Kruijssen2014} and would be interesting for higher-resolution follow-ups, e.g., with the main ALMA 12m array.

\begin{figure*}
    \centering%
    \includegraphics[width=\textwidth]{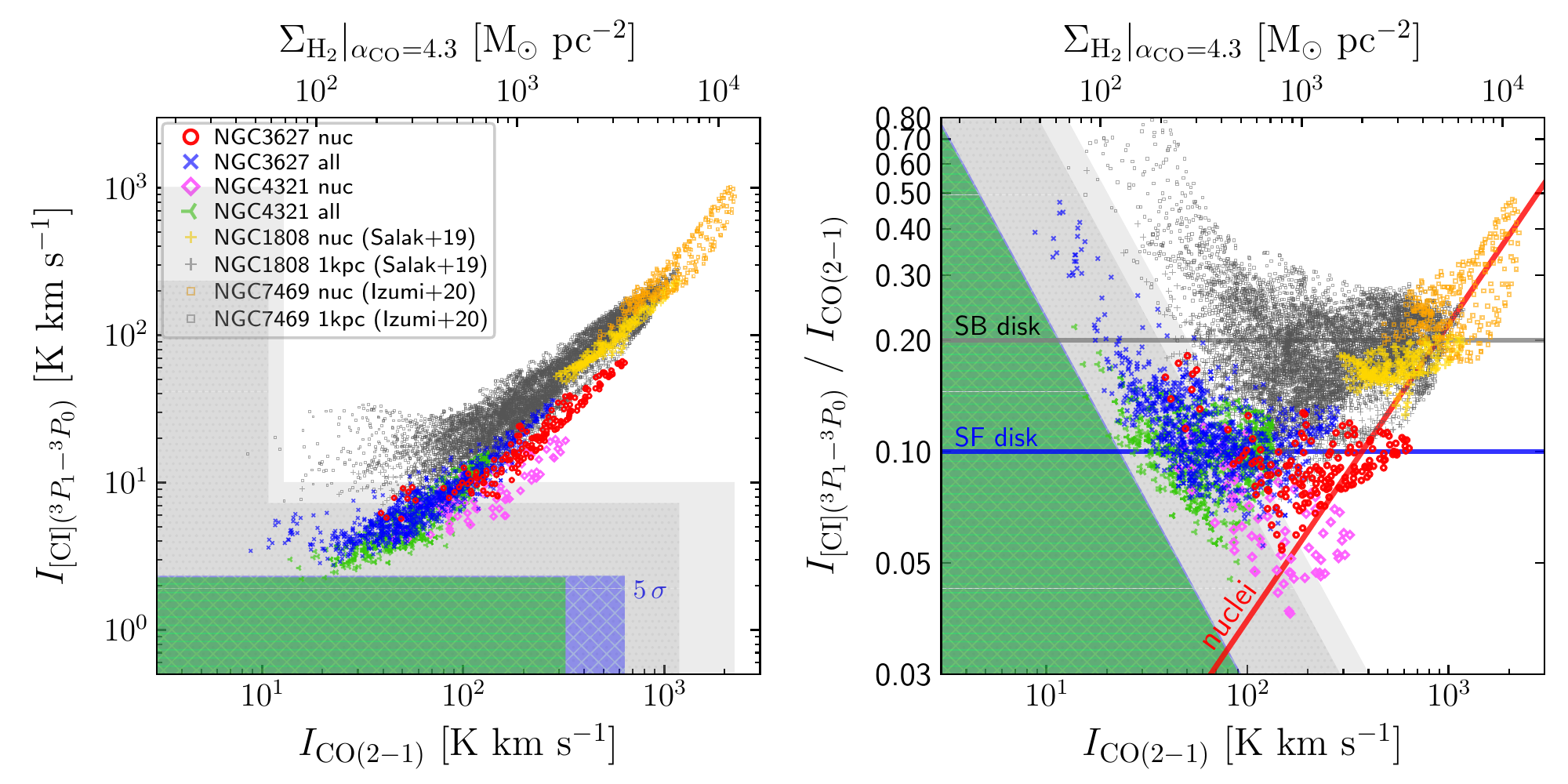}
    \caption{%
        \textit{Left}: $\Remark{ICI10}$ versus $\Remark{ICO21}$ in the four galaxies we studied: NGC~3627, NGC~4321, NGC~1808 and NGC~7469. $\Remark{ICI10}$ and $\Remark{ICO21}$ integrated fluxes are moment-0 intensities extracted from their beam and pixel size matched data cubes. Only data points with propagated moment-0 $\SNR \ge 5$ are shown. 
        The blue (with crossed hatch), green (with horizontal line hatch), gray (with dotted hatch) and light gray shadings indicate the $5\,\sigma$ thresholds for the four galaxies, respectively. Typical line widths of 10~km~s$^{-1}$ (NGC~3627 and NGC~4321) and 30~km~s$^{-1}$ (NGC~1808 and NGC~7469) in sigma are used to calculate the thresholds together with the channel rms in Table~\ref{table: targets}. 
        For NGC~3627 and NGC~4321, their ($uv$-matched) CO(2--1) $5\,\sigma$ thresholds are out of the plotting range.
        Top axis shows the simply-converted H$_2$ surface density using a constant $\alphaCO = 4.3$ and CO excitation (i.e., CO(2--1) to CO(1--0) line ratio) $R_{21} = 0.65$. 
        \textit{Right}: Same data but shown as the line ratio $\RCICO = \Remark{ICI10}/\Remark{ICO21}$ versus $\Remark{ICO21}$. 
        The blue horizontal line indicates a star-forming (SF) disk-like $\RCICO \sim 0.1$, i.e., the spiral arms of NGC~3627 and NGC~4321. The gray horizontal line indicates a starburst (SB) disk-like $\RCICO \sim 0.2$, i.e., the $\sim 1$~kpc SB disk/ring in NGC~1808 and NGC~7469. 
        The red line with a slope of 0.8 in logarithm ($0.035 \, (I_{\Remark{CO21}}/[100\,\Kkms])^{0.8}$) visually guides the trend of increasing $\RCICO$ in the central regions of these galaxies (see boxes in Figs.~\ref{fig: NGC3627 line ratio map} to \ref{fig: other line ratio maps}, sizes $\sim 300$--500~pc). 
    }
    \label{fig: line ratio vs CO}
\end{figure*}

\begin{figure*}
    \centering%
    \includegraphics[width=0.493\textwidth]{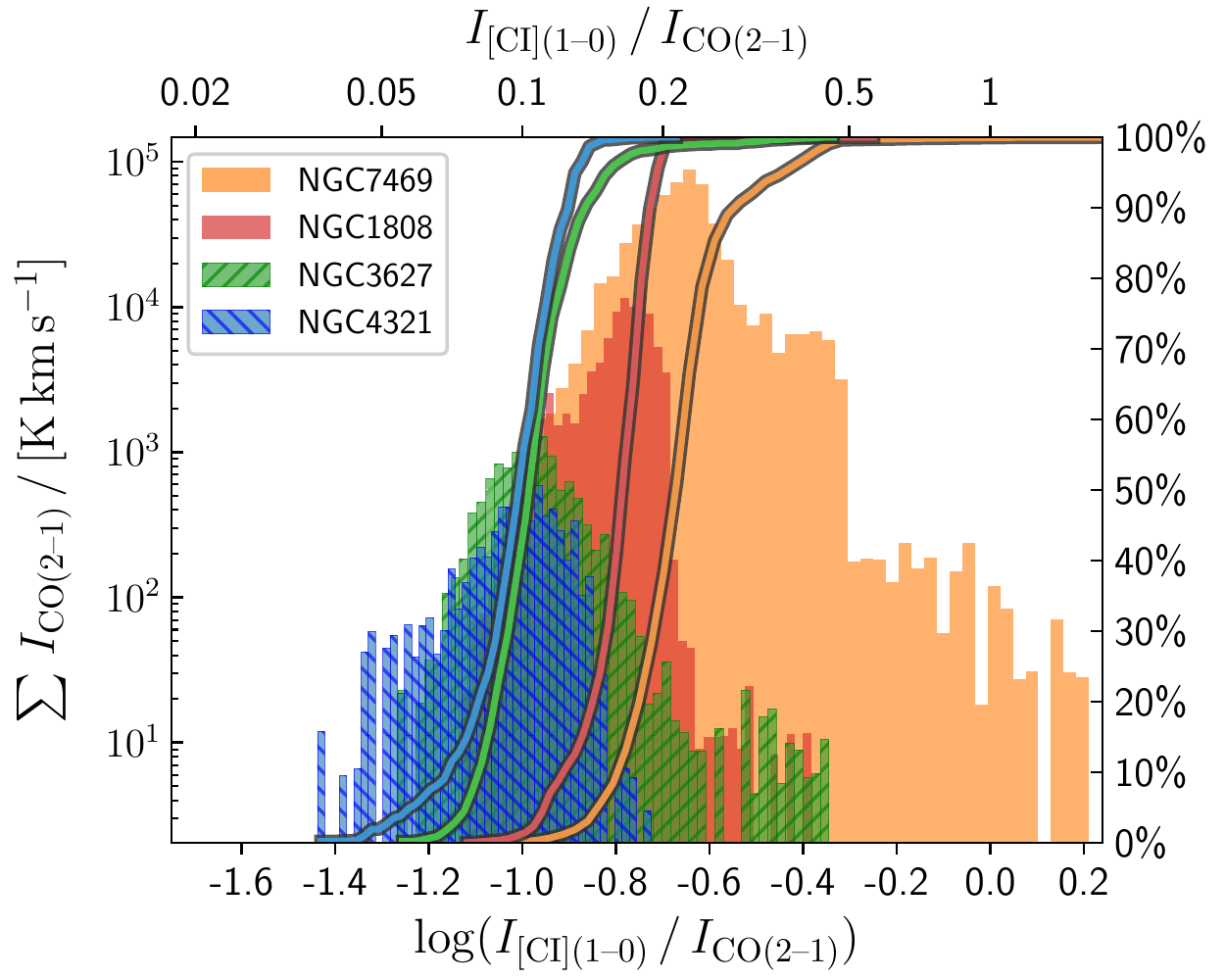}
    \includegraphics[width=0.493\textwidth]{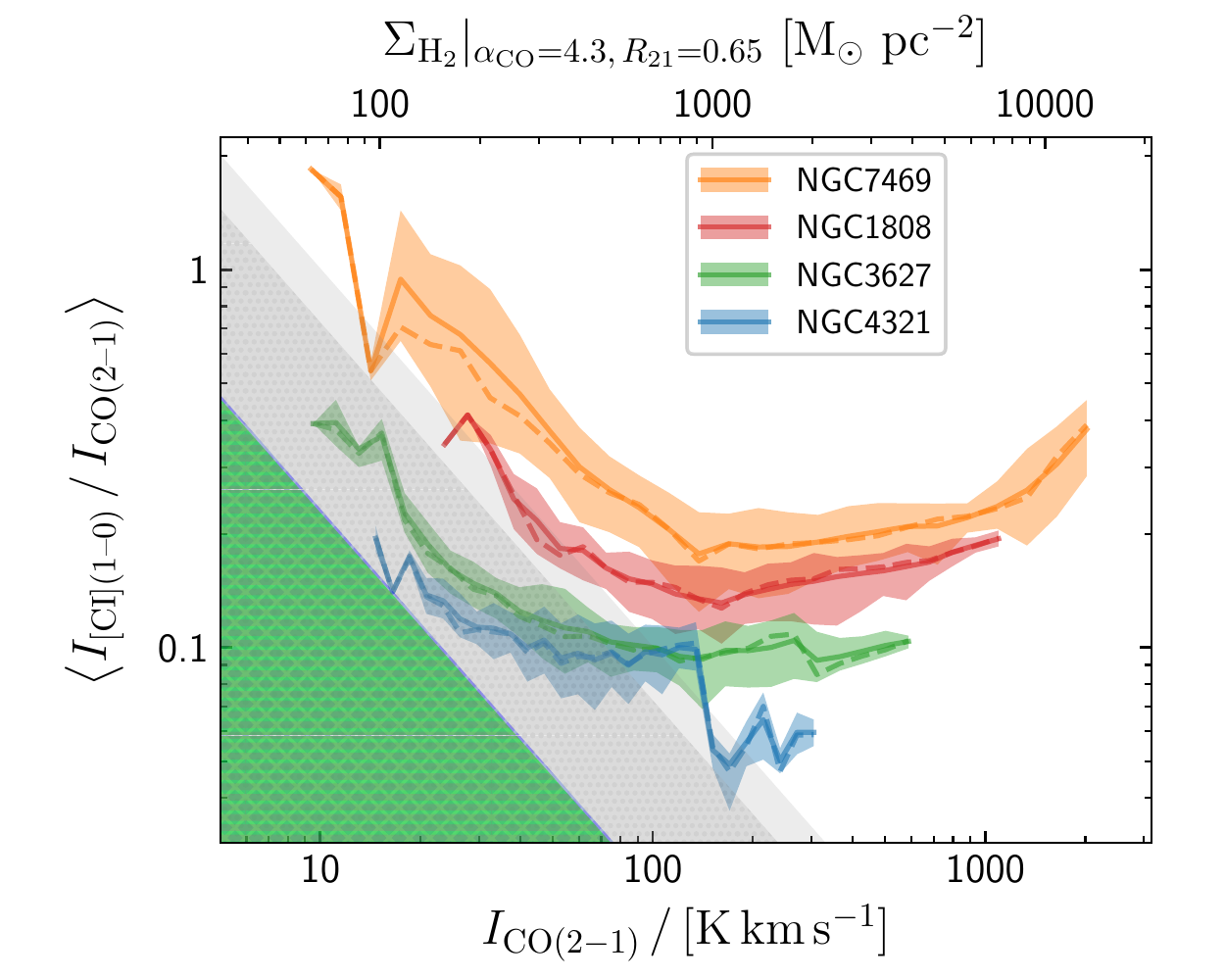}
    \caption{%
    \textit{Left panel} shows the histograms of the sum of CO(2--1) surface brightness over pixels in each \CI/\CO{} line ratio interval in the four galaxies. The bold curves are the cumulative distributions of the corresponding histograms. 
    \textit{Right panel} shows the mean \CI/\CO{} line ratio for pixels in each bin of CO(2--1) surface brightness (or molecular gas surface density as indicated by the top-axis assuming a constant $\alphaCO = 4.3$ and $R_{21} = 0.65$). Shaded area enveloping each mean trend indicates the 16- to 84-th percentiles in each bin, and dashed line indicates the 50-th percentile. 
    This panel thus more clearly illustrates the ridge and scatter of the trends in the right panel of Fig.~\ref{fig: line ratio vs CO}. 
    Detection limits are shown consistently with Fig.~\ref{fig: line ratio vs CO}. 
    }
    \label{fig: line ratio histogram}
\end{figure*}

\subsection{\RCICO\ versus CO surface brightness and gas surface density}

In Fig.~\ref{fig: line ratio vs CO} we show the scatter points of \CI{} and CO line brightnesses and the \RCICO{} line ratios. 
Systematic differences in $\RCICO$ can be found among these galaxies and within their galactic environments. We discuss these environments separately as follows.

\subsubsection{Star-forming and starburst disk environments}

The majority of molecular gas in the 3--7~kpc star-forming disks of NGC~3627 and NGC~4321 exhibits a tight distribution of $\RCICO$ around 0.1 (within about $\pm$50\%). The distribution is fairly flat over an order of magnitude in $\Remark{ICO21}$. We denote this as the ``SF disk'' regime.
The ratio rises to about 0.2 within a galactocentric radius of few hundred parsec to $\sim 1$~kpc starburst areas in NGC~1808 and NGC~7469, which we define as the ``SB disk'' regime. 

The nearly flat ``SB disk'' distribution is consistent with earlier studies. \citet{Salak2019} found that in NGC~1808 $\Remark{ICI10}$ is proportional to $\Remark{ICO21}^{\ +1.05}$. They also studied CO(1--0), finding that $\Remark{ICI10} \propto \Remark{ICO10}^{\ +1.46}$, i.e., with a steeper slope. 
\citet{Saito2020} found a similar slope between $\Remark{ICI10}$ and $\Remark{ICO10}$: $\Remark{ICI10} \propto \Remark{ICO10}^{\ +1.54}$ in IRAS~F18293-3413 (whose CO(2--1) data is insufficient for such a study as previously mentioned). 
This correlation slope is likely to change with different CO transitions due to the CO excitation's dependence on the CO luminosity or star formation rate themselves (e.g., \citealt{Daddi2015,Liudz2021a}). 
For higher-$J$ CO transitions, \citet{Michiyama2021} studied \CI{} and CO(4--3) lines in 36 (U)LIRGs, finding $\Remark{ICI10} \propto \Remark{ICO43}^{\ +0.97}$. 

The upward tails seen at the CO-faint-end in the right panel of Fig.~\ref{fig: line ratio vs CO} are mostly due to the detection limits in each galaxy. The color shading indicates the 5-$\sigma$ limit for each galaxy, assuming a typical line width of 10~km~s$^{-1}$. Region~22 of NGC~3627 and the outer pixels in the two starburst/AGN galaxies make the tips of the blue and gray faint-end tails, respectively. But most of these features should still be spurious. We performed CASA simulations following the real observations of NGC~3627 to verify what we would get if \RCICO{} were constant across the galaxies (Appendix~\ref{appendix: simulation})
We find that noise can indeed boost a simulated \RCICO{} of 0.2 to a measurement as high as 0.5 in a NGC~3627-like large mosaic, or even higher to $>1$ in single-pointing ALMA data without total power or sufficient short spacing. However, the faint-end tail from the simulation is not very similar to that of NGC~3627, raising some questions about the nature of the high-\RCICO{} in Region~22. 

Furthermore, we also did stacking experiments using the CO line profile as the prior to measure the \CI{} line fluxes in bins of CO line brightnesses or galactocentric radii. We find no systematically higher \RCICO{} at fainter CO pixels or at the outer radii.

\subsubsection{Galaxy centre environments}

Most interestingly, the $\lesssim 250$~pc centres of all four galaxies show dramatically different $\RCICO$. It varies by a factor of ten from $\sim 0.05$ in the centre of NGC~4321  to $\sim 0.1$ in the centre of NGC~3627, then to $\sim 0.15$--0.2 at the centre of NGC~1808, and to $\sim 0.2$--0.5 in the nucleus of NGC~7469. 
This ``nuclei'' trend has a slope of about 0.8, i.e., $\Remark{ICI10} \approx 0.035 \; \Remark{ICO21}^{\ +0.8}$, and corresponds to a $\Sigmol$ range of about $10^{3}$ to $10^{4} \; \Mspc2$ if adopting an $\alphaCO = 4.3 \ \mathrm{M_{\odot} \; (K \, km \, s^{-1} \, pc^2)^{-1}}$ (\citealt{Bolatto2013}) and a CO excitation $R_{21} = 0.65$ (\citealt{Leroy2021c}). 

The reasons for the nuclei showing this trend likely include both the effect of varying [\CI/CO] abundance ratio and the excitation and radiative transfer. They are determined by the actual ISM conditions in each environment, i.e., the gas kinetic temperature, \CI{} and CO column densities ($\CIcolumndensity$ and $\COcolumndensity$), line widths ($\Delta{v}$), and $\mathrm{H_2}$ volume density ($\nHtwo$), etc. 
They will be analyzed in Sect.~\ref{sec: Excitation Analysis}.

\subsubsection{Histogram and curve-of-growth of the line ratio distributions}

Fig.~\ref{fig: line ratio histogram} shows the histogram and cumulative distribution functions of $\RCICO$ (left panel) and the mean $\RCICO$ and the 16-th to 84-th percentiles in bins of CO line intensity (right panel) in our galaxies. 
The histogram is computed as the sum of CO line intensity in each bin of $\RCICO$, representing the amount of gas at each $\RCICO$. 
The NGC~4321 histogram is lower than others, showing that it has the least total molecular gas mass among the four galaxies within mapped areas. The NGC~3627 histogram has a peak about 0.5~dex higher than that of NGC~4321, but their peak locations are consistently at $\log \RCICO \sim -1$, same conclusion as our previous scatter plots show. 
The two SB galaxies NGC~1808 and NGC~7469, despite having much smaller observed areas ($\sim 1$--$2$~kpc), have more than an order of magnitude higher peaks hence total molecular gas masses. 
The locations of their histogram peaks also shift to a systematically higher $\RCICO$. 
It is clearly seen that both SF and SB disks lack a high-$\RCICO$ ($\log \RCICO > -0.35$) gas component which dominates the gas in the strong AGN host galaxy NGC~7469.

The curve-of-growth in the left panel of Fig.~\ref{fig: line ratio histogram} further shows the cumulative CO brightness in bins of $\RCICO$, and reflects the fact that the majority of gas in our galactic disks has a narrow distribution of $\RCICO$ (see also last few rows in Table~\ref{table: targets}). 
In NGC~3627, NGC~4321 and NGC~1808, the higher- and lower-$\RCICO$ gas outside $\pm 0.1$~dex around the mean $\RCICO$ only sums up to $<5$--$10\%$ of their total gas masses. 

In the right panel of Fig.~\ref{fig: line ratio histogram}, we illustrate how the mean $\RCICO$ changes with different gas surface density, as well as the scatter of the data point distribution. The 2D pixels are binned by CO brightness, and the mean and 16\%/50\%/84\%-th percentiles of the $\RCICO$ are computed for each bin. 
It sketches out the mean trends in Fig.~\ref{fig: line ratio vs CO} (right panel), and also reveals that 
a) the scatter of $\RCICO$ is the largest ($\pm 0.20$~dex) in the starburst disk of NGC~7469 compared to other galaxies; 
b) the scatter is the smallest ($\pm 0.06$~dex) in the star-forming disks of NGC~4321 and NGC~3627 where $\Sigmol \sim 100 \text{--} 500 \; \Mspc2$; 
c) the scatters are very similarly ($\pm 0.10$~dex) at a relatively high $\Sigmol \sim 1000 \; \Mspc2$ in all our galaxies except for NGC~4321's centre.

\subsection{\RCICO\ versus ISRF}
\label{subsection: ISRF}

\begin{figure*}
\centering%
\includegraphics[width=0.9\textwidth]{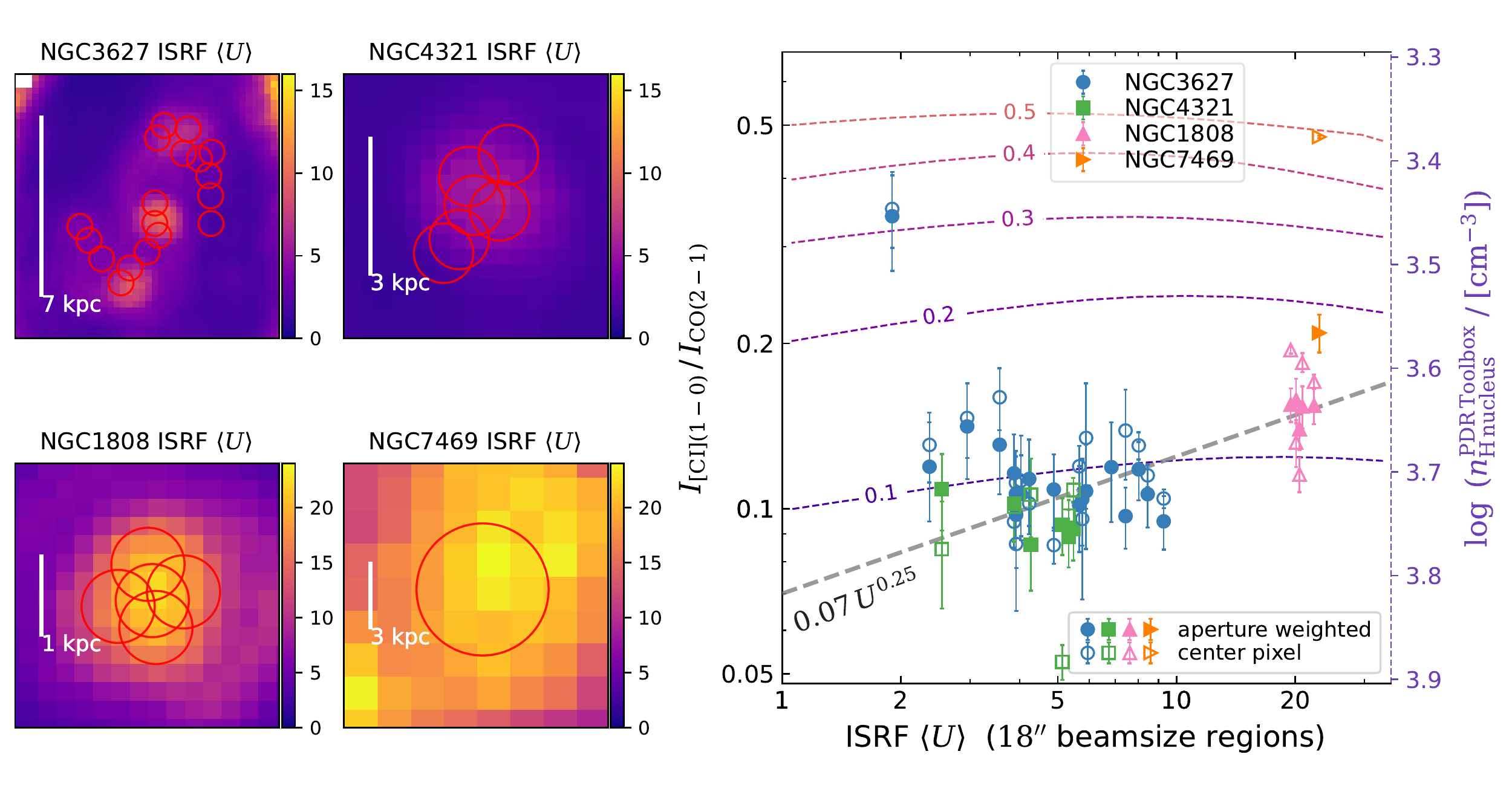}
\caption{%
    Line ratio versus mean ISRF intensity $\Umean$ measured from SED fitting to the near- to far-IR (\textit{Spitzer}, \textit{Herschel}) data. 
    \textit{Left four panels} are the $\Umean$ maps of NGC~3627, NGC~4321, NGC~1808 and NGC~7469, and as described in Sect.~\ref{subsection: ISRF} are obtained from the works done by J. Chastenet et al. (in prep.) as a continued effort of \citet{Chastenet2021}. 
    \textit{Right panel} shows the scatter plot of $\RCICO$ versus $\Umean$ at a spatial resolution of $\sim 18''$ in the manually-selected apertures shown in the left panels. 
    We computed a beam-weighted mean $\RCICO$ and a centre pixel $\RCICO$ for each aperture, displayed as the solid and open symbols, respectively. 
    Error bars are the corresponding beam-weighted mean or central pixel \RCICO{} uncertainties in each region. 
    The region diameter ($\sim 18''$) corresponds to the coarsest resolution of the \textit{Herschel} data used in the SED fitting. 
    The gray dotted line shows $\RCICO = 0.07 \, U^{0.25}$ as a guiding line. 
    The colored dotted contours (labeled 0.1--0.5) are the model contours of $\RCICO$ computed from the PDR Toolbox (using the latest \incode{wk2020} model and solar metallicity; \citealt{Kaufman2006}). We convert $G_0$ from the model contour to the $x$-axis $U$ by multiplying with a factor of 0.88 (\citealt{Draine2007}). 
    The right $y$-axis indicates the PDR model's $\nHtwo$, whose range is manually adjusted so that the model contours roughly match the left $x$-axis. 
    The PDR model contours show little dependence of $\RCICO$ on $U$ in the horizontal direction but a strong correlation with $\nHtwo$ as seen by the vertical gradient. 
\label{fig: line ratio vs ISRF}
}\vspace{1ex}
\end{figure*}

Here we study how \RCICO{} correlates with the interstellar radiation field (ISRF) intensity ($U$; in units of Milky Way mean ISRF; \citealt{Habing1968}), which is a measure of the UV radiation field strength impacting the ISM. Since \CI{} originates from PDRs, its abundance should be highly related to the PDR properties, thus whether or how \RCICO{} correlates with the ISRF that illuminates the PDR is a key question. 

The ISRF intensity $U$ is usually measured by the re-radiated dust emission, as the original UV photons are highly attenuated by dust in the ISM. \citet[][hereafter DL07]{DL07} developed a series of dust grain models and synthesized SED templates that can describe nearby galaxies' observed dust SEDs. A number of studies (e.g., \citealt{Draine2007,Galliano2011,Dale2012,Aniano2012,Aniano2020,Daddi2015,Chastenet2017,Liudz2021a,Chastenet2021}) have used the DL07 models to fit the dust SEDs of galaxies near and far. 

For our study, we adopt the mass-averaged ISRF $\Umean$ maps from an ongoing effort by J. Chastenet et al. (in prep.) which is a continuation of the work published for other galaxies by \citet{Chastenet2017,Chastenet2021} with the same method. 
In this method, the \textit{Herschel} far-infrared, \textit{Spitzer} and/or \textit{WISE} near-infrared images\,\footnote{For the two (U)LIRGs, the WISE maps are saturated, so their inferred ISRF intensity $U$ should be considered with a large uncertainty.} are first PSF-matched to a common resolution $\sim 18''$ (\textit{Herschel} SPIRE 250$\mu$m), then SED fitting is performed using the DL07 models at each pixel. 
Each dust SED is fitted by two dust components, one in the ambient ISM with a constant ISRF intensity $U_{\mathrm{min}}$, and the other in regions with higher $U$ with a power law distributed ISRF from $U_{\mathrm{min}}$ up to $U_{\mathrm{max}} = 10^{7}$.
The fitting then gives the best-fit $U_{\mathrm{min}}$ and the power-law index $\gamma$ of the $U$ distribution. 
The mean ISRF intensity $\Umean$ is calculated as the dust mass-weighted mean of the ambient and power-law ISRF intensities (in DL07, $U_{\rm PDR} \ge 10^2$ is considered to be associated with the PDR), and better represents the overall ISRF than $U_{\mathrm{min}}$. 

Fig.~\ref{fig: line ratio vs ISRF} presents the $\Umean$ maps and the \RCICO{} versus $\Umean$ scatter plot. Since the spatial resolution of the $\Umean$ maps is $18''$, we calculate a weighted $\RCICO$ in each circular aperture shown in Fig.~\ref{fig: line ratio vs ISRF} whose diameter equals the $\Umean$ map resolution, with weighting following the 2D Gaussian beam.
In addition, we also show the central pixel value for each region in Fig.~\ref{fig: line ratio vs ISRF}. 
We restrict the analysis to regions where both $\Umean$ and \RCICO{} are measured. 

A weak correlation between \RCICO{} and $\Umean$ is seen from the SF to SB disks, with a large scatter. For reference, we plot a $\RCICO = 0.07 \, U^{0.25}$ line to indicate this weak trend. The lowest-$\Umean$ data point in NGC~3627 (mainly Region~22) and the centre of NGC~7469 do not follow this trend, possibly indicating other \CI{} enhancement mechanisms than $\Umean$. 

We use the popular PDR modelling software, PDR Toolbox (version 2.2.9; \citealt{Kaufman2006,Pound2008,pdrtoolbox})\,\footnote{\url{https://dustem.astro.umd.edu/}; \url{https://github.com/mpound/pdrtpy}.}
to compute the theoretical $\RCICO$ in uniform grids of ISRF intensity in the \citet{Habing1968} $G_0$ unit and H nucleus density $n_{\mathrm{H}}$ (both in log space). We adopt the latest \incode{wk2020} model set and a solar metallicity ($\mathrm{[O/H_2]} = 3.2 \times 10^{-4}$, $\mathrm{[C/H_2]} = 1.6 \times 10^{-4}$). 
The ISRF intensity grid can be directly scaled to $U$ by multiplying a factor 0.88 (\citealt{Draine2007}), corresponding to the $x$-axis of Fig.~\ref{fig: line ratio vs ISRF}. 
The $n_{\mathrm{H}}$ grid is not directly linked to the left $y$-axis of Fig.~\ref{fig: line ratio vs ISRF}, but because the model contours show fairly flat $\RCICO$ versus $U$ trends, we can match the model contours and indicate the $n_{\mathrm{H}}$ grid in the right $y$-axis of Fig.~\ref{fig: line ratio vs ISRF}. 

The PDR modelling predicts generally flat $\RCICO$ versus $U$ trends similar to the distribution of the observed data points, although our $U$ are diluted values at a much larger physical scale than in the PDR models. 
However, in PDR models, \RCICO{} changes moderately with $n_{\mathrm{H}}$, and a relatively high $n_{\mathrm{H}} \sim 10^{3.5\text{--}3.7} \, \mathrm{cm^{-3}}$ is needed to reproduce an $\RCICO \sim 0.1$--0.2, or equivalently $\sim 1.0$--1.9 for the ratio of $\Remark{CI10}$ and $\Remark{CO21}$ fluxes in $\mathrm{erg\,s^{-1}\,cm^{-2}\,sr^{-1}}$ units used by the PDR Toolbox\,\footnote{See \url{https://github.com/mpound/pdrtpy-nb/blob/master/notebooks/PDRT_Example_Model_Plotting.ipynb}. 
We scale the ratio of intensities in $\mathrm{erg\,s^{-1}\,cm^{-2}\,sr^{-1}}$ to the ratio of intensities in $\mathrm{K\,km\,s^{-1}}$ by a factor of $(\nu_{\mathrm{CI10}}/\nu_{\mathrm{CO21}})^{-3} = 9.73$, where $\nu$ is the rest-frame frequency of the lines. We caution that the old version of \incode{PDRToolbox} \incode{v2.1.1} does not include the \incode{CI_609/CO_21} line ratio, but has \incode{CI_609/CO_43} and \incode{CO_43/CO_21} ratios. However, we tested that the multiplication of \incode{CI_609/CO_43} and \incode{CO_43/CO_21} does not produce the same result as the direct \incode{CI_609/CO_21} ratio in the new \incode{PDRToolbox} \incode{v2.2.9}. We use the new version for this work.
}. 
Furthermore, the PDR model needs a $\sim0.5$~dex lower $n_{\mathrm{H}}$ to explain the $2\times$ elevated \RCICO{} in SB disks, which seems somewhat counterintuitive. 

This apparent tension is likely caused by the observations probing a much coarser resolution that consists of a large number of PDRs with different $n_{\mathrm{H}}$ and $U$ (e.g., increasing $U$ from 10 to 400 can increase $\RCICO$ from 0.2 to 0.5 based on our PDR Toolbox modelling). It may also relate to an abundance variation driven by non-PDR mechanisms --- 
X-ray dominated region (XDR; \citealt{Maloney1996,Meijerink2005,Meijerink2007}; see also review by \citealt{Wolfire2022}) and Cosmic Ray Dominated Region (CRDR; \citealt{Papadopoulos2010c,Papadopoulos2011,Papadopoulos2018,Bisbas2015,Bisbas2017}) are well-known to play an important role in starbursting and AGN environments, respectively.

\begin{figure*}[htb]
\centering
\includegraphics[width=0.9\textwidth]{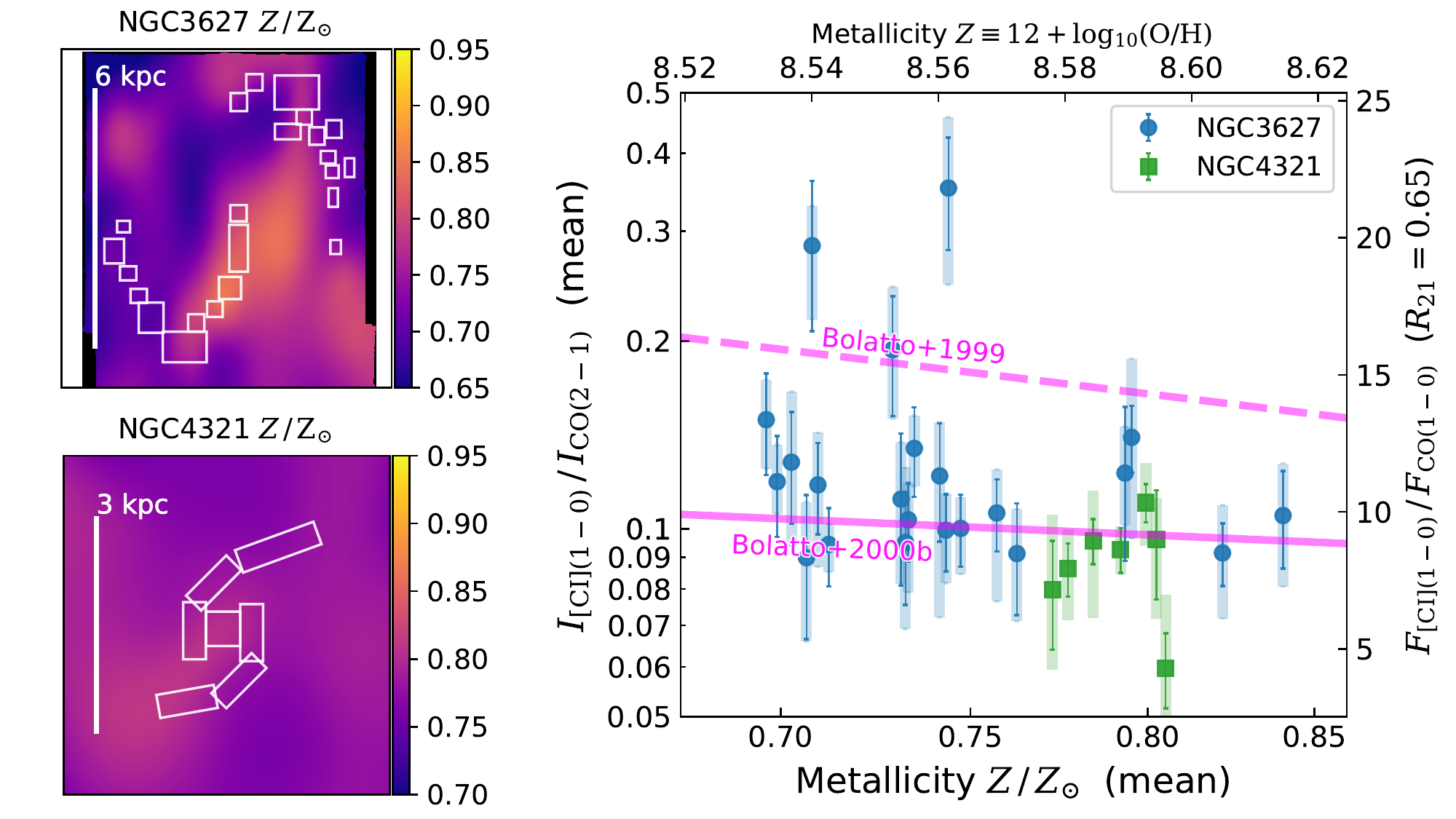}
\caption{%
    \textit{Left two panels} show the metallicity maps of NGC~3627 and NGC~4321 from \citet{Williams2022} based on PHANGS-MUSE (\citealt{PHANGSMUSE}) data. Boxes are regions as defined in Figs.~\ref{fig: NGC3627 line ratio map} and \ref{fig: NGC4321 line ratio map}. 
    \textit{Right panel} shows the \RCICO{} versus metallicity scatter plot for regions in NGC~3627 and NGC~4321. The bottom-axis is metallicity in solar units, top-axis is metallicity in $12+\log(\mathrm{O/H})$, left-axis is the line ratio of $\Remark{CI10}$ and $\Remark{CO21}$ in $\Kkms$ units, and right-axis is the ratio of fluxes in $\mathrm{erg\,s^{-1}\,cm^{-2}\,sr^{-1}}$ units. 
    The left- and right-axes are matched by a factor of $(\nu_{\mathrm{C{\textsc{i}}1\textnormal{--}0}}/\nu_{\mathrm{CO1\textnormal{--}0}})^3 \times R_{21}$, where we adopt a $R_{21}=0.65$ (\citealt{Leroy2021c}). 
    Data points are the mean metallicity and \RCICO{} in each region shown in left panels, with error bars indicating the typical uncertainty in \RCICO{}, and vertical color shading indicating the 16- to 84-th percentiles of pixels' \RCICO{} distribution inside each region. 
    The \citet{Bolatto2000b} empirical fitting and \citet{Bolatto1999} model prediction lines are overlaid. 
    The three outliers in NGC~3627 with $\RCICO \sim 0.19$, 0.28 and 0.35 are Regions 20, 21 and 22, respectively. The outlier point in NGC~4321 with $\RCICO \sim 0.06$ is its centre Region 1.
    \label{fig: line ratio vs metallicity}
}\vspace{1ex}
\end{figure*}

As pointed out by \citet{Papadopoulos2011} and later works, ultra-luminous infrared galaxies (ULIRGs) with IR luminosities $L_{\mathrm{IR,\,8-1000\mu\mathrm{m}}} \ge 10^{12} \, \Lsun$ and corresponding $\mathrm{SFRs} > 1000 \, \Msyr$ have cosmic-ray (CR) energy densities $\sim 10^{3} \text{--} 10^{4}\times$ larger than the Milky Way. In our sample, the two SB galaxies are not as intensively star-forming as ULIRGs, but their SFRs are still a factor of 2 to 20 higher than our SF galaxies. More importantly, the SB disks have considerably higher (e.g., one to two orders of magnitude) SFR surface densities than the SF disks, plausibly leading to a factor of a few tens to hundreds higher CR intensities. The CRs emitted from OB star clusters and supernova remnants can penetrate deeply into the ISM, and thus nearly uniformly illuminate the ISM and effectively dissociate CO into \CI{}. 

A detailed simulation by \citet{Bisbas2017} showed that $10\times$ and $100\times$ higher CR ionization rates ($\xi_{\mathrm{CR}}$) than the Milky Way level results into a factor of about 2 and 3 mildly-increased \CI{} abundance and about 1/2 and 1/30 strongly-decreased CO abundance, respectively (see their Fig.~3), in a 10~pc GMC with a mean gas density $\left< n_{\mathrm{H}} \right> = 760 \, \mathrm{cm^{-3}}$, resembling Milky Way and typical SF galaxies' ISM conditions. 
Our SB galaxies probably have a $\xi_{\mathrm{CR}}$ around or slightly above 10, but not reaching 100 given their mild $\Umean \sim 20-30$ compared to the $\Umean \gtrsim 50$ in extreme SB galaxies, e.g., local ULIRGs and high-redshift hyperluminous IR-bright (HyLIRG) and submm galaxies (\citealt{Daddi2015,Silverman2018,Liudz2021a}). 
In Sect.~\ref{sec: Excitation Analysis}, we show that the mildly-increased [\CI/CO] abundance ratio from a CRDR and temperature/density-driven excitation can lead to the factor of 2 increase in \RCICO{} when going from SF to SB disks.

\subsection{\RCICO\ versus Metallicity}
\label{subsec: metallicity}

Our current sample covers a relatively small range in metallicity for a metallicity--line ratio study. Given their stellar masses and metallicity maps (e.g., \citealt{Kreckel2019,Kreckel2020,Williams2022}), our targets probe a near-solar metallicity. NGC~3627 has a well-measured %(negative) 
metallicity gradient of $Z = 8.53 + 0.06 \times [r/R_{25}]$ from \citet{Kreckel2019}\,\footnote{\citet{Kreckel2019} used a distance of $10.6 \pm 0.9 \; \mathrm{Mpc}$ for NGC~3627. We do not directly use this metallicity gradient, but use the metallicity maps from \citet{Williams2022} where the same new distance as in this work is adopted.}, but the metallicity of individual \ionizedhydrogen{} regions scatter between $\sim 8.45$--$8.65$. 
\citet{Williams2022} produced the 2D spatial distributions of gas-phase metallicity in PHANGS-MUSE galaxies, including NGC~3627 and NGC~4321, by applying the Gaussian Process Regression to map the smooth, higher-order metallicity variation from the individual, sparse \ionizedhydrogen{} regions detected in PHANGS-MUSE (\citealt{PHANGSMUSE, Santoro2021}). 
This method is significantly superior to a na\"{i}ve nearest neighbor interpolation method. We refer the reader to \citet{Williams2022} for the technical details and validation demonstration. 

In Fig.~\ref{fig: line ratio vs metallicity} we present the trend between \RCICO{} and metallicity in NGC~3627 and NGC~4321. 
Due to our limited metallicity range, only a very weak decreasing trend can be seen. 
A similar trend has been reported by \citet{Bolatto2000b}, who observed CO(1--0) and $\Remark{CI10}$ in the Region N27 of the Small Magellanic Cloud and combined with literature data to present a correlation between \RCICO{} and metallicity over $\sim 0.2 \text{--} 2 \, Z_{\odot}$. 
We overlay the \citet{Bolatto2000b} empirical fitting\,\footnote{%
We adopt the equation from \citep[][Fig.~2 caption]{Bolatto2000b}, with a typo-correction in the brackets so that the equations are:
$\log ( F_{\Remark{CI10}} / F_{\Remark{CO10}} ) = -0.47 \, \left[ 12 + \log \mathrm{(O/H)} \right] + 5.0$ and $\log ( F_{\Remark{CI10}} / F_{\Remark{CO10}} ) = -0.82 \, \left[ 12 + \log \mathrm{(O/H)} \right] + 8.2$, respectively, where $F$ is the line flux in $\mathrm{erg \, s^{-1} \, cm^{-2} \, sr^{-1}}$. 
}
as well as a theoretical PDR model prediction line from \citet{Bolatto1999} in Fig.~\ref{fig: line ratio vs metallicity}. 
Although our data covers only a small metallicity range, it is encouraging to see the overall good agreement with the observationally-driven trend in \citet{Bolatto2000b}. 
More detailed understanding of the deviation from PDR models will likely need more \CI{} mapping in lower-metallicity galaxies.

\vspace{2ex}

\section{Estimation of abundance ratios under representative ISM conditions}
\label{sec: Excitation Analysis}

The observed \RCICO{} depends on not only the \CI{} and CO abundances (\COabundance{} and \CIabundance{}), but also the ISM conditions that determine the excitation (level population) and radiative transfer of the C atoms and CO molecules --- i.e., the gas kinetic temperature $\Tkin$, volume density of H$_2$ ($\nHtwo$; as the primary collision partners of C and CO), turbulent line width $\dv$, and the species' column densities $\COcolumndensity$ and $\CIcolumndensity$. 
The column density divided by the line width ($N/\dv$) further determines the optical depth $\tau$ of each line. 

In this work, having only a CO(2--1) and a $\Remark{CI10}$ line is insufficient to numerically solve all the $\Tkin$, $\nHtwo$, $\COcolumndensity/\dv$, \COabundance{} and \CIabundance{}. Therefore, we assume several \textit{representative ISM conditions} where $\Tkin$, $\dv$ and \COabundance{} are fixed to reasonable values, then solve the $\COcolumndensity$ and $\CIcolumndensity$. We assume that \CI{} and CO are spatially mixed at our resolution, i.e., they have the same filling factor. This is a relatively strong assumption and is likely not the real case, because although we see spatial co-localization at our $\sim 200$~pc resolutions, the spectral profiles of CO and \CI\ show discrepancies (Appendix~\ref{appendix: all regions}), indicating different underlying distributions at smaller scales. With the assumption of the same filling factor (and 3D distribution), the column density ratio $\CIcolumndensity/\COcolumndensity$ equals the [\CI/CO] abundance ratio. 

We perform both LTE and non-LTE calculations. 
LTE assumes that all transitions are thermalized locally, and the level population follows Boltzmann statistics, so that each transition's excitation temperature $T_{\mathrm{ex}}$ as defined by the level population equals $\Tkin$ (e.g., \citealt{Spitzer1998Book,Tielens2010Book,Draine2011Book}). The LTE assumption facilitates the computation as we do not need to solve the level population and thus $\nHtwo$ is not used. 
However, LTE is not easily reached in typical ISM conditions. Subthermalized lines need non-LTE calculations. The commonly adopted non-LTE approach is the Large Velocity Gradient (LVG; \citealt{Scoville1974,Goldreich1974}), where an escape fraction is used to facilitate the local radiation intensity calculation hence the detailed balance of radiative and collisional (de-)excitations. 
We use RADEX (\citealt{RADEX}) to compute the non-LTE \CI{} and CO line fluxes for each fixed parameter set ($\Tkin$, $\nHtwo$, $\COcolumndensity$, $\CIcolumndensity$, $\dv$), then study how [\CI/CO] determines the line flux ratio $\RCICO$. 

We assume the following 7 representative ISM conditions:
\textit{(1)} inner Galactic GMCs, 
\textit{(2)} NGC~3627/4321 disk GMCs, 
\textit{(3)} NGC~3627 bar-end environments, 
\textit{(4)} NGC~1808/7469 SB disk, 
\textit{(5)} NGC~4321 centre, 
\textit{(6)} NGC~3627 centre, 
and \textit{(7)} NGC~7469 centre. 
Conditions \textit{(1)}--\textit{(3)} correspond to the ``SF disk'' in Fig.~\ref{fig: line ratio vs CO}, and condition \textit{(4)} represents the ``SB disk''. 
The last three conditions are for the ``nuclei''. 
Under each ISM condition, we set a fixed $\Tkin$ (in units of K), $\dv$ (in units of km~s$^{-1}$), \COabundance{}, and $\Sigmol$ (in units of $\Mspc2$). 
The adopted values are shown at the top of each panel in Figs.~\ref{fig: LTE RCICO} and \ref{fig: NonLTE RCICO}. 

Our choice of $\Tkin$ considered multi-line measurements in the literature, e.g., \citet{Heyer2015} for Galactic GMC ISM condition, and \citet{Teng2022} for NGC~3627 nucleus, as well as typical values in simulations (e.g., \citealt{Offner2014,Glover2015,Glover2016,Clark2019,HuChiaYu2021a}). 
Our $\dv$ is based on the observed line width in the highest-resolution ALMA data, i.e., from the original PHANGS-ALMA $1''$ data (\citealt{Leroy2021phangs}) or from \citet{Salak2019} and \citet{Izumi2020} for NGC~1808 and NGC~7469, respectively. The observed line width should at least provide an upper limit, if not directly probing the turbulent $\dv$. 
The \COabundance{} is more like a fiducial value and does not obviously affect later [\CI/CO] results. 
We also take a fiducial $\Sigmol$ which matches the observed CO(1--0) line brightness temperature and a CO-to-H$_2$ conversion factor $\alphaCO \sim 4 \text{--} 0.8 \; \mathrm{M_{\odot} \, (K \, km \, s^{-1})^{-1}}$ depending on the assumed conditions (e.g., \citealt{Bolatto2013}; and our paper~II). 
This $\Sigmol$ directly converts to an $N_{\mathrm{H_2}}$ if assuming that 36\% of the mass is helium. Then $N_{\mathrm{H_2}}$ and \COabundance{} together indicate a \COcolumndensity{}. Only with these assumptions can we unambiguously link an abundance ratio [\CI/CO] (i.e., $\CIcolumndensity/\COcolumndensity$) to a line flux ratio $\RCICO{}$. 

Additionally, with the same filling factor assumption, $\RCICO{}$ is also equivalent to the ratio of the CO-to-H$_2$ and \CI{}-to-H$_2$ conversion factors, i.e., $\RCICO = \alphaCO/\alphaCI$, because they trace the same molecular gas at our resolution. 
In a companion paper (paper II; Liu et al. 2022 in prep.), we study the behaviors of $\alphaCO$ and $\alphaCI$ in more details, also under both LTE and non-LTE conditions.

\subsection{LTE results}
\label{subsec: LTE}

\begin{figure*}[htb]
    \centering
    \includegraphics[width=0.99\textwidth]{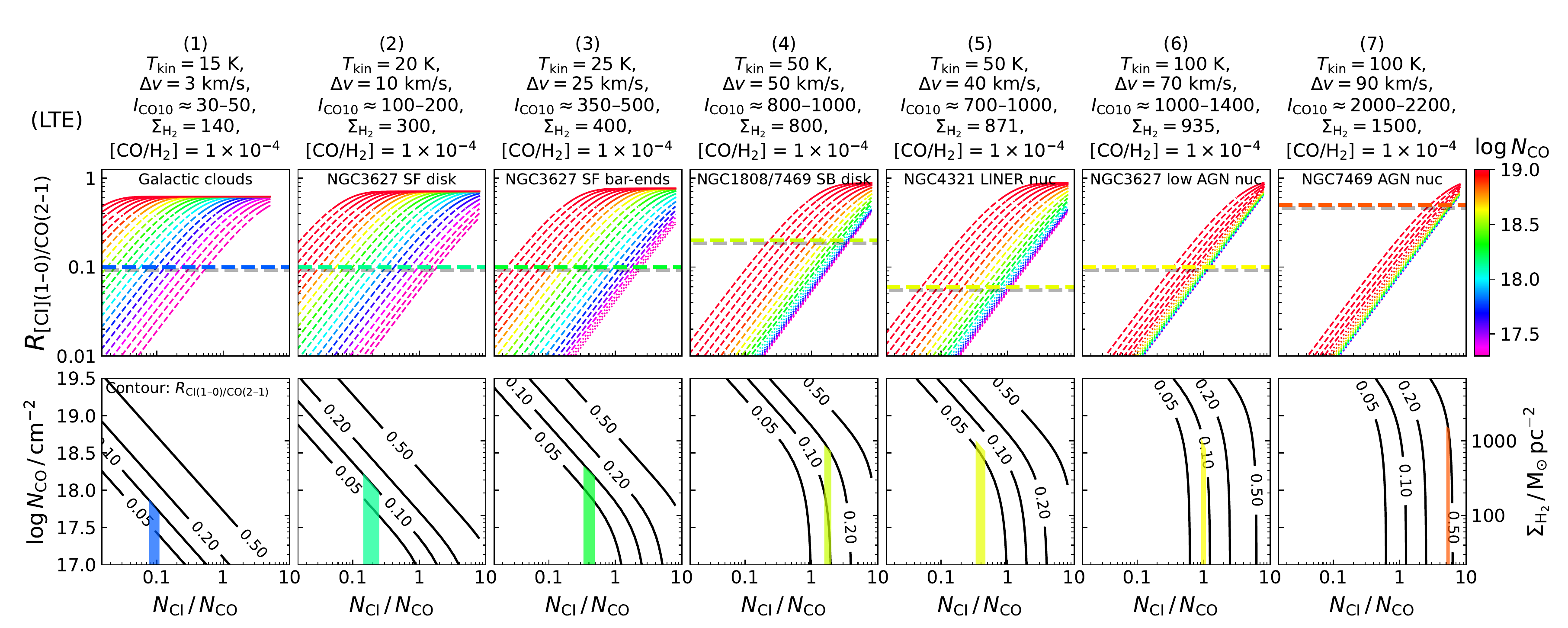}
    \caption{%
        Observed/calculated $\RCICO$ as a function of CO column density \COcolumndensity{} and [\CI/CO] abundance (column density) ratio $\RNCINCO$ for 7 representative ISM conditions, based on LTE calculations. 
        \textit{Upper panels}: Curved lines are $\RCICO$ versus $\RNCINCO$ for different \COcolumndensity{} as indicated by the color. Dashed means CO(2--1) optically thick and $\Remark{CI10}$ optically thin, dotted means both are both optically thin, and solid line means both optically thick. 
        Horizontal thick line is the observed $\RCICO$ value. 
        \textit{Lower panels}: Black contours are $\RCICO = 0.05, \, 0.10, \, 0.20$ and $0.50$ in the CO column density versus \CI/CO abundance ratio plane. 
        Vertical color shading indicates the final $\RNCINCO$ that can match both the observed $\RCICO{}$ and the assumed $\Sigmol$ value noted at the top of each panel.
    }
    \label{fig: LTE RCICO}
\end{figure*}

We determine the LTE solution of [\CI/CO] in Fig.~\ref{fig: LTE RCICO}. 
To illustrate the complexity of this problem and the degeneracy of the $\Tkin$, $\COcolumndensity/\dv$ and [\CI/CO] parameters, we show how $\RCICO$ ($y$-axis) varies with $\RNCINCO$ ($x$-axis) and \COcolumndensity{} (color bar) in the upper panels, and how $\RCICO$ (contour) can be pinpointed by determining the $\Sigmol$ (right $y$-axis). 
At the top of each column the key parameters of the aforementioned representative ISM conditions are shown as text. The assumed $\Tkin$ and $\dv$ generally increase from panels (1) to (7) mimicking real conditions. Observational-driven $\Remark{ICO10}$ ranges and the assumed $\Sigmol$ for these ISM environments are also listed, and can be converted one into the other via $\alphaCO$. We also list the chosen \COabundance{} for clarity.

As shown in the upper panels, $\RCICO$ increases with both $\RNCINCO$ and \COcolumndensity{} under low-$\Tkin$ conditions. However, a maximum $\RCICO$ is reached in the high \CI{} column density regime, where an increasing abundance ratio will no longer increase $\RCICO$. The line ratio saturates at about 0.5 when $\Tkin \sim 15$~K, or at about 1 for $\Tkin \sim 100$~K. The observed line ratios (horizontal dashed lines) are all lower than the saturation limit of $\RCICO{}_{\mathrm{max}} \sim 0.7 \text{--} 0.9$, and corresponds to a range of $\COcolumndensity$ or $\Sigmol$. This range is very wide in the low-$\Tkin$ conditions, with $\log \COcolumndensity / \mathrm{cm^{-2}} \sim 17 - 19$, but is very narrow at $\Tkin \sim 100$~K, that is, $\RCICO$ can be determined from [\CI/CO] without much dependence on $\COcolumndensity$. 

In the lower panels, the contours of $\RCICO = 0.05$, 0.1, 0.2 and 0.5 are shown in the plane of CO column density $\COcolumndensity$ versus \CI/CO abundance ratio. 
For a fixed $\RCICO{}$, a higher $\RNCINCO$ corresponds to a lower $\COcolumndensity$ in the low-$\Tkin$ regime, but \RCICO{} becomes insensitive to $\COcolumndensity$ at a high $\Tkin$. In the latter case, $\RNCINCO$ alone determines $\RCICO{}$. 
By matching to the expected $\Sigmol$ for each ISM condition, we can obtain an unambiguous $\RNCINCO$ from the observed \RCICO{} as shown in Fig.~\ref{fig: LTE RCICO}. 

In this way, we determine the $\RNCINCO$ to be around 0.1--0.2 for conditions \textit{(1)} and \textit{(2)}, and $\sim 0.4$ for conditions \textit{(3)} and \textit{(5)}. 
The SB disk (condition \textit{4}) and AGN environments (conditions \textit{6}/\textit{7}) have $\RNCINCO \sim 2$ and $\sim 1 \text{--} 10$, respectively. 

The strongly enhanced \CI/CO abundance ratios in SB disk and AGN environments are clear signatures of CRDR and XDR, respectively. 
The NGC~4321 centre is a low-ionization nuclear emission-line region (LINER, e.g., \citealt{GarciaBurillo2005}), and given our assumptions it requires $\RNCINCO \sim 0.4$ in order to explain the observed low line ratio. 
The gradually increasing \CI/CO abundance ratios in the nuclei of NGC~4321, NGC~3627, NGC~1808 and NGC~7469 agree well with the increasing power of AGN activities, e.g., their X-ray luminosities (Table~\ref{table: targets}).

Our assumptions for the Galactic cloud ISM condition are mainly based on \citet{Heyer2009}, \citet{Heyer2015} and \cite{RomanDuval2010}. The obtained [\CI/CO]~$\sim 0.1$ also agrees well with previous studies using \CI{} and CO isotopologues (see Sect.~\ref{subsec: intro RCICO}. 
For example, \citet{Ikeda1999} reported $\RNCINCO \sim 0.05 \textnormal{--} 0.2$ for the Orion cloud; 
\citet{Oka2001} found $\RNCINCO = 0.064 \pm 0.035$ but a peak of $\RNCINCO \sim 0.15$ in the DR 15 \HII{} region and two IR dark clouds; 
\citet{Kramer2004} found $\RNCINCO \sim 0.11 \textnormal{--} 0.12$ in the massive SF region W3; 
\citet{Genzel1988} found $\RNCINCO \sim 0.2$ for the massive SF region W51; 
\citet{Zmuidzinas1988} reported $\RNCINCO \sim 0.05 \textnormal{--} 0.15$ in several dense clouds. 
Recently, \citet{Izumi2021} derived an overall $\RNCINCO \sim 0.1$ in the massive SF region RCW38 at $A_V \sim 10 \textnormal{--} 100$, consistent with the Orion cloud, but also saw high [\CI/CO] $\sim 0.2 \textnormal{--} 0.6$ in some even lower column density regions. 

Accurate determination of all the ISM properties instead of simple assumptions as in this work requires comprehensive, multi-$J$ CO isotopologue observations (e.g., \citealt{Teng2022}). 
Our representative ISM conditions only provide a qualitative diagnostic to understand how to infer the \CI/CO abundance ratio from the observed \RCICO{} quantity. 
Their properties may not be very accurate, and are subject to change with future multi-tracer/CO isotopologue studies.

\begin{figure*}[htb]
\centering
\includegraphics[width=\textwidth]{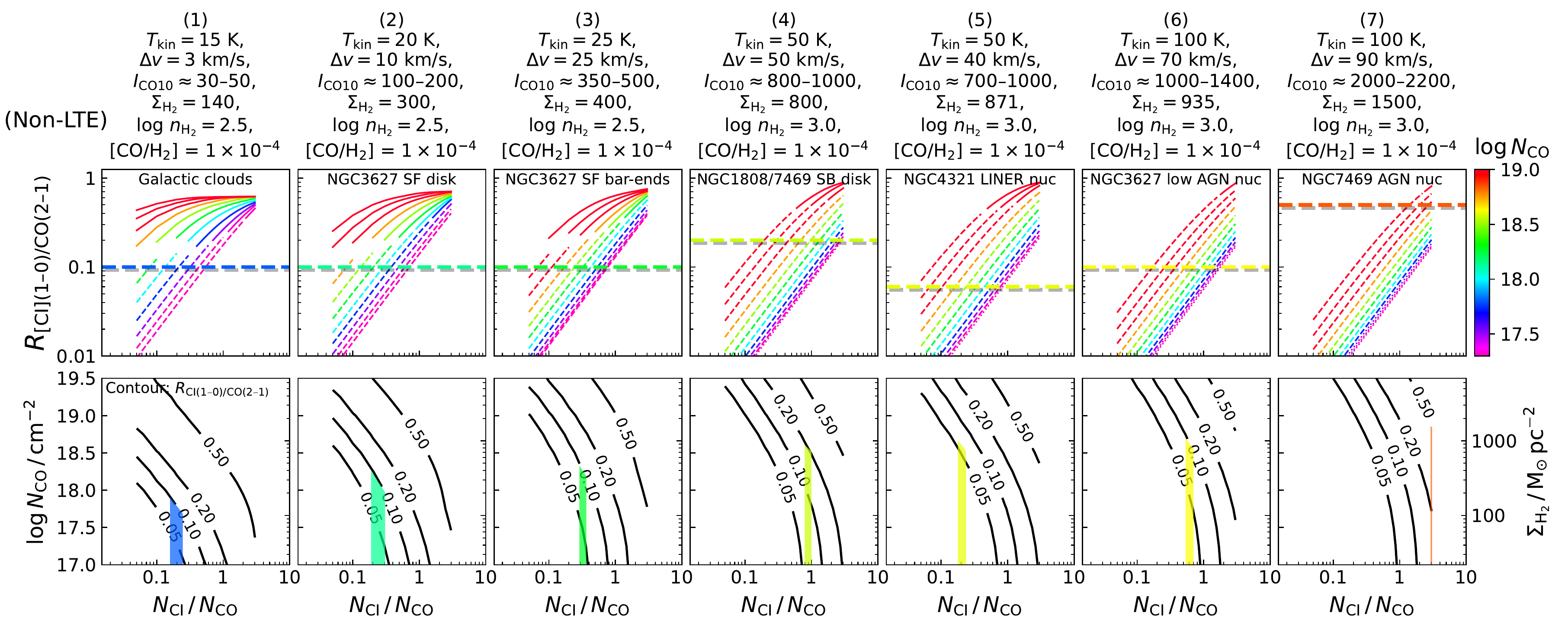}
\caption{%
Similar to Fig.~\ref{fig: LTE RCICO} but with non-LTE calculations.
See Fig.~\ref{fig: LTE RCICO} caption for the details. 
}
\label{fig: NonLTE RCICO}
\end{figure*}

\subsection{Non-LTE results}
\label{subsec: non-LTE}

We perform the non-LTE calculations using the RADEX code (\citealt{vanderTak2007})
with the Large Velocity Gradient (LVG; i.e., expanding sphere) escape probability (\citealt{Scoville1974,Goldreich1974}). 
Comparing to LTE, the volume density of H$_2$ is now an additional property that plays a role. The excitation temperature of each line no longer equals $\Tkin$. 
We compute the non-LTE line ratios in a grid of $( \Tkin, \, \dv, \, \COabundance, \, \RNCINCO, \, \COcolumndensity, \, \nHtwo )$ parameter sets, then study the line and abundance ratios similarly as in the previous section. 

In Fig.~\ref{fig: NonLTE RCICO} we show the non-LTE version of Fig.~\ref{fig: LTE RCICO}. 
The non-LTE result does not qualitatively differ from that of LTE.
The most obvious difference is that at high temperature ($\Tkin > 50$~K, given our other assumed conditions), the required [\CI/CO] is lower by $\times 2$ for the same observed $\RCICO$, whereas at low temperature ($\Tkin \sim 15$~K) a $\lesssim 2 \times$ [\CI/CO] is expected for the observed line ratio. 

As listed in the title of each panel, we assume $\log \nHtwo / \mathrm{cm^{-3}} = 2.5$ to 3.0 for our conditions (1) to (7). 
This is based on the fact that low-$J$ CO and \CI{} trace the bulk of molecular gas with $\log \nHtwo / \mathrm{cm^{-3}} \sim 2$--3 (e.g., \citealt{Leroy2017,Liudz2021a}), with a typical value $\nHtwo \sim 300 \, \mathrm{cm^{-3}}$ widely adopted/found for Galactic and nearby galaxy GMCs (e.g., \citealt{Koda2012,Pety2017}, see also simulations of \citealt{Glover2012b}), whereas a smaller portion of their emission is from the dense gas with $\log \nHtwo / \mathrm{cm^{-3}} \sim 4$--6 (e.g., \citealt{Gao2004a}). 
The fraction of the dense gas is expected to increase in more starbursting environment (\citealt{Gao2004b}), so does the mean gas density (\citealt{Liudz2021a}). Therefore, a $\log \nHtwo / \mathrm{cm^{-3}}$ of 3.0 is perhaps a reasonable assumption, or at least a lower limit for our galaxy centres. A $10 \times$ higher density does not qualitatively change Fig.~\ref{fig: NonLTE RCICO} except for shifting the expected $\RNCINCO$ to be 50\% higher in the $\Tkin > 50$~K conditions (making Fig.~\ref{fig: NonLTE RCICO} more like Fig.~\ref{fig: LTE RCICO}). 

Therefore, our non-LTE result strongly confirms the LTE result in the previous section, that the intrinsic \CI/CO abundance has to vary significantly from about 0.1--0.2 to about 1--3, in order to explain the observed line ratios from Galactic GMCs (hundred pc scales) and NGC~3627/4321 SF disk GMCs to NGC~1808/7469 SB disk/AGN ($r \lesssim 250$~pc) environments.

\vspace{2ex}

\section{Discussion}
\label{sec: Discussion}

From this study we see that the line ratio $\RCICO$ increases from 0.1 to 0.2 and $> 0.5$ from the SF disk to the SB disk and the vicinity (few hundred pc) of X-ray luminous AGN. Although we may be limited by our sample, 
we suggest that this trend can be extended to general cases at both local and high redshift, because the \CI/CO abundance ratio enhancement is well-predicted by the CRDR and XDR theories. 

\citet{Papadopoulos2010c} and \citet{Papadopoulos2011} pointed out the key role of CRDR in local (U)LIRGs. With UV photons alone, cold gas in starbursting environments can still be well shielded, hence as cold as $\sim 10$~K. However, with CRs produced from O, B star clusters and supernova remnants, the gas can be uniformly illuminated and heated up. 
The extremely starbursty local ULIRGs have a CR density ($\xi_{\mathrm{CR}}$) $10^{3}$--$10^{4}$ times that of the Milky Way (\citealt{Papadopoulos2010c}), and thus will naturally dissociate more CO into \CI{}. 
\citet{Bisbas2017} and \citet{Papadopoulos2018} conducted simulations of molecular gas at tens of pc scale with varying CR strength, and found that when $\xi_{\mathrm{CR}}$ is $\sim 1 \textnormal{--} 100 \, \times$ Galactic value, the gas temperature in the dense gas remains largely unaffected ($\sim 10$~K). 
However, when $\xi_{\mathrm{CR}}$ reaches $\sim 1000 \, \times$ Galactic value as in (U)LIRGs, the gas temperature increases to 30--50~K. The \citet{Bisbas2017} simulation also shows that the \CI{} and \ionizedcarbon{} fractional abundances at $\log A_V \sim 0.5$ increase by one and three orders of magnitude, respectively, and the CO fractional abundance decreases correspondingly by over two orders of magnitude, with an increasing $\xi_{\mathrm{CR}}$ from the Galactic value to $100 \, \times$ of that. 
These results are in line with our calculations in Sect.~\ref{subsec: LTE}. 

Meanwhile, the AGN-driven XDR chemistry and molecule/atom excitation have been firstly modeled in great detail by \citet{Maloney1996}, \citet{Meijerink2005} and \citet{Meijerink2007}. In XDRs, photo-ionization becomes the dominant heating source instead of photo-electric emission from dust grains in PDRs, thus more molecules are dissociated and/or ionized. \citet{Meijerink2005}'s Model~1 shows that CO molecules are fully dissociated when $N_{\mathrm{H}} < 10^{23} \, \mathrm{cm^{-2}}$, then the fractional abundance is gradually increased by six orders of magnitudes from $N_{\mathrm{H}} = 10^{23} \, \mathrm{cm^{-2}}$ to $10^{24} \, \mathrm{cm^{-2}}$. By contrast, the \CI{} and \ionizedcarbon{} abundances are about constant until reaching the densest part ($N_{\mathrm{H}} > 10^{24.5} \, \mathrm{cm^{-2}}$). Unlike in PDRs, there is no clear \ionizedcarbon{}/\CI{}/CO stratification in XDRs (\citealt{Meijerink2007,Wolfire2022}). Their Model~2 with a $100 \times$ higher energy deposition rate shows that the CO distribution contracts to the highest gas density spots, and the total cumulative [\CI/CO] is increased by about 0.2~dex, reaching $\RNCINCO \sim 1.6$. 
This value agrees well with our calculations shown in Figs.~\ref{fig: LTE RCICO} and \ref{fig: NonLTE RCICO} (the ISM condition 7). 

Therefore, both in theoretical and observational views, it is clear that the $\RCICO$ line ratio is a good indicator of the starburst's CRDR and AGN's XDR. 
If a $\RCICO$ line ratio (to be specific, $\Remark{ICI10}/\Remark{ICO21}$) reaches 0.2--0.5, then the CRDR plays a crucial role and the abundance ratio [\CI/CO] could be $\sim 1$.
If it exceeds 0.5--1.0, then it may only be enhanced by an XDR, e.g., caused by an X-ray luminous AGN, which boosts the abundance ratio [\CI/CO] to $\gtrsim 1$--2.

\vspace{2ex}

\section{Conclusion}
\label{sec: Conclusion}

We presented new beam- and $uv$-coverage-matched $\Remark{CI10}$ and CO(2--1) maps in the galactocentric radius $r \lesssim 7$~kpc SF disk of NGC~3627 at $\sim 190$~pc resolution, and in the $r \lesssim 3$~kpc disk of NGC~4321 at $\sim 270$~pc resolution. 
They are among the largest \CI{} mosaic mappings available in nearby galaxies. 

We combined the imaging of $\Remark{CI10}$ and CO(2--1) in two nearby, more starbursty and AGN-hosting galaxies, i.e., NGC~1808 central $\sim1$~kpc, and NGC~7469 central $\sim1.5$~kpc at 140--160~pc resolutions. 
Together, we studied the spatial distributions of \CI{} and CO, and their line ratio $\RCICO$ as functions of various galaxy resolved properties, including CO brightness (Fig.~\ref{fig: line ratio vs CO}), ISRF $\Umean$ (Fig.~\ref{fig: line ratio vs ISRF}), and metallicity (Fig.~\ref{fig: line ratio vs metallicity}). 

In order to obtain the underlying intrinsic [\CI/CO] abundance ratio from the observed line ratio $\RCICO$, we assumed 7 representative ISM conditions and did both LTE and non-LTE calculations. 

\vspace{1ex}
We summarize our findings as below: 
\begin{itemize}[itemsep=0ex,topsep=1ex,leftmargin=4ex]
    \item The majority of the molecular gas in the SF disks of NGC~3627 and NGC~4321 has a uniform $\Remark{RCI10CO21} = 0.10 \pm 0.05$. 
    $\Remark{CI10}$ and $\Remark{CO21}$ exhibit very similar spatial distributions at our 140--270~pc resolution.
    
    \item The majority of the molecular gas in the SB disk of NGC~1808 has $\Remark{RCI10CO21} = 0.20 \pm 0.05$. 
    This elevated line ratio compared to the SF disk is likely caused by a $\sim 3 \text{--} 5 \times$ increased [\CI/CO] abundance ratio, based on our LTE and non-LTE calculations and parameter assumptions (Figs.~\ref{fig: LTE RCICO} and \ref{fig: NonLTE RCICO}). 
    
    \item The centres ($r \lesssim 250$~pc) of NGC~4321, NGC~3627, NGC~1808 and NGC~7469 show a strongly increasing trend in $\Remark{RCI10CO21}$, from 0.05 to 0.5, as a function of $I_{\Remark{CO21}}$ --- i.e., $\Remark{RCI10CO21} \approx 0.035 \, (I_{\Remark{CO21}}/[100\;\Kkms])^{0.8}$. 
    The inferred [\CI/CO] abundance ratio from our non-LTE calculation changes from $\sim 0.2$ to $\sim 3$, in line with the increasing X-ray luminosity of the centres/AGNs. 
    
    \item The observed $\RCICO$ versus ISRF trend is fairly flat, i.e., \RCICO{} is insensitive to ISRF intensity changes, similar to the PDR model prediction calculated with PDRToolbox (although our resolution is much coarser than the physical scale of PDR models; Sect.~\ref{subsection: ISRF}). 
    
    \item We find a $\RCICO$ versus metallicity trend consistent with previous observations that covered lower-metallicity environments ($\sim 0.2 \, \mathrm{Z_{\odot}}$; \citealt{Bolatto2000b}), within the metallicity range ($\sim 0.7 \text{--} 1 \, \mathrm{Z_{\odot}}$) of our sample (Sect.~\ref{subsec: metallicity}). 
    
    \item Our inferred [\CI/CO] abundance ratios agree well with the CRDR and XDR's theoretical studies. 
    Given the drastically changed \RCICO{}, we propose that the \CI{} to CO line ratio can be a promising indicator of CRDR (starbursting) and XDR (AGN). For instance, an observed line flux ratio of
    $\Remark{RCI10CO21} \sim 0.2 \textnormal{--} 0.5$ and $\gtrsim 0.5$ effectively indicates CRDR and XDR, respectively. 
    
    \item We do not find ubiquitous ``CO-dark'', \CI{}-bright gas at the outer $r\sim 1 \text{--} 3$~kpc disks of NGC~3627 and NGC4321, nor systematic enhancement in \RCICO{} at CO-faintest sightlines via stacking, mainly due to the incompleteness of our data.
    Nevertheless, a few sightlines, e.g., Region~22 in NGC~3627 show high \RCICO{}~$\sim 0.5$ which do not fully comply with the noise behavior seen in our CASA simulations.
    
\end{itemize}

\vspace{2ex}

\begin{acknowledgements}
We thank the anonymous referee for very helpful comments. 
ES, TS and TGW acknowledge funding from the European Research Council (ERC) under the European Union’s Horizon 2020 research and innovation programme (grant agreement No. 694343).
ER acknowledges the support of the Natural Sciences and Engineering Research Council of Canada (NSERC), funding reference number RGPIN-2017-03987.
The work of AKL is partially supported by the National Science Foundation under Grants No. 1615105, 1615109, and 1653300.
AU acknowledges support from the Spanish grants PGC2018-094671-B-I00, funded by MCIN/AEI/10.13039/501100011033 and by ``ERDF A way of making Europe'', and PID2019-108765GB-I00, funded by MCIN/AEI/10.13039/501100011033. 
RSK and SCOG acknowledge financial support from the German Research Foundation (DFG) via the collaborative research centre (SFB 881, Project-ID 138713538) “The Milky Way System” (subprojects A1, B1, B2, and B8). They also acknowledge funding from the Heidelberg Cluster of Excellence ``STRUCTURES'' in the framework of Germany’s Excellence Strategy (grant EXC-2181/1, Project-ID 390900948) and from the European Research Council via the ERC Synergy Grant ``ECOGAL'' (grant 855130). 
FB and IB acknowledge funding from the European Research Council (ERC) under the European Union’s Horizon 2020 research and innovation programme (grant agreement No.726384/Empire).
JC acknowledges funding from the European Research Council (ERC) under the European Union’s Horizon 2020 research and innovation programme DustOrigin (ERC-2019-StG-851622).
MC gratefully acknowledges funding from the Deutsche Forschungsgemeinschaft (DFG) in the form of an Emmy Noether Research Group (grant number CH2137/1-1). 
Y.G.’s work is partially supported by National Key Basic Research and Development Program of China (grant No. 2017YFA0402704), National Natural Science Foundation of China (NSFC, Nos.
12033004, and 11861131007), and Chinese Academy of Sciences Key Research
Program of Frontier Sciences (grant No. QYZDJ-SSW-SLH008).
AH was supported by the Programme National Cosmology et Galaxies (PNCG) of CNRS/INSU with INP and IN2P3, co-funded by CEA and CNES, and by the Programme National “Physique et Chimie du Milieu Interstellaire” (PCMI) of CNRS/INSU with INC/INP co-funded by CEA and CNES.
K.K.\ gratefully acknowledges funding from the German Research Foundation (DFG) in the form of an Emmy Noether Research Group (grant number KR4598/2-1, PI Kreckel)
JMDK and MC gratefully acknowledge funding from the Deutsche Forschungsgemeinschaft (DFG) in the form of an Emmy Noether Research Group (grant number KR4801/1-1) and the DFG Sachbeihilfe (grant number KR4801/2-1), and from the European Research Council (ERC) under the European Union’s Horizon 2020 research and innovation programme via the ERC Starting Grant MUSTANG (grant agreement number 714907).
JP acknowledges support from the Programme National “Physique et Chimie du Milieu Interstellaire” (PCMI) of CNRS/INSU with INC/INP co-funded by CEA and CNES.
HAP acknowledges support by the Ministry of Science and Technology of Taiwan under grant 110-2112-M-032-020-MY3.
The work of JS is partially supported by the Natural Sciences and Engineering Research Council of Canada (NSERC) through the Canadian Institute for Theoretical Astrophysics (CITA) National Fellowship.

This paper makes use of the following ALMA data: 
\incode{ADS/JAO.ALMA#2015.1.00902.S}, 
\incode{ADS/JAO.ALMA#2015.1.00956.S}, 
\incode{ADS/JAO.ALMA#2015.1.01191.S}, 
\incode{ADS/JAO.ALMA#2017.1.00078.S}, 
\incode{ADS/JAO.ALMA#2017.1.00984.S}, 
\incode{ADS/JAO.ALMA#2018.1.00994.S}, 
\incode{ADS/JAO.ALMA#2018.1.01290.S}, 
\incode{ADS/JAO.ALMA#2019.1.01635.S}.

ALMA is a partnership of ESO (representing its member states), NSF (USA) and NINS (Japan), together with NRC (Canada), MOST and ASIAA (Taiwan), and KASI (Republic of Korea), in cooperation with the Republic of Chile. The Joint ALMA Observatory is operated by ESO, AUI/NRAO and NAOJ.

\end{acknowledgements}

\bibliographystyle{aa_url} % enabling URLs in References, from https://github.com/yangcht/AA-bibstyle-with-hyperlink/blob/master/aa_url.bst
%\bibliography{Biblio.bib}

\begin{appendix}

\section{Data reduction, imaging and moment map generation}
\label{appendix: data reduction}

After downloading the raw data, i.e., 12m+7m for $\Remark{CO21}$ and 7m for $\Remark{CI10}$, respectively, we ran the observatory calibration pipeline (``scriptForPI'') to reduce the data. 
As shown in \citet{Leroy2021pipeline}, the observatory calibration produces well-calibrated visibility data without further manual flagging. 
We then split out the visibility data and proceeded to the imaging step. 

In order to match the spatial resolution of $\Remark{CO21}$ and $\Remark{CI10}$, we carefully matched their $uv$ ranges. 
We flagged the $\Remark{CO21}$ visibility data outside the $uv$ range of 19--68~m ($\sim$15--52k$\lambda$) so as to match the $uv$ range of the $\Remark{CI10}$ data (minimum to 80\% percentile: 8.9--32~m, or $\sim$15--52k$\lambda$). 
Given that the $\Remark{CI10}$ $uv$ plane distribution is elliptic, while the $uv$ range selection in \incode{CASA} is circular, our chosen thresholds are adjusted to make the final $\Remark{CO21}$ synthesized beam similar to that of $\Remark{CI10}$. 
Note that in a later step we also convolved both image cubes to the exact same rounded beam ($3.46''$) before generating the moment maps. 

The \textit{PHANGS-ALMA imaging \& post-processing pipeline}\,\footnote{\url{https://github.com/akleroy/phangs_imaging_scripts}; version 2.0.} developed by \citet{Leroy2021pipeline} is used for creating line cubes and moment maps. The code is suitable for general nearby galaxy strong line mapping interferometric and total power mosaic data. In brief, the visibility data are carefully transformed into the user-given channel width and frequency range corresponding to the line plus line-free channels, then continuum is subtracted by fitting an order of zero polynomial. Then, two stages of imaging (``cleaning'') are performed: first a multi-scale cleaning down to $\sim4\sigma$, then a single-scale cleaning down to $\sim1\sigma$. At each stage, \incode{CASA}'s \incode{tclean} task is iteratively called so as to force many major cycles, which has proven to improve the accuracy of the deconvolution. Before and during the second stage cleaning, a clean mask is computed at each iteration according to the outcome of previous iteration by finding all $>2\sigma$ pixels associated with $>4\sigma$ peaks in a watershed algorithm (\citealt{Rosolowsky2006,Rosolowsky2021}). In this way the cleaning can include as much signal as possible and ensure a signal-free residual. 

Then the image cubes are corrected for the primary beam attenuation, converted to Kelvin units, and convolved to a common beam, which is the coarser beam of $\Remark{CO21}$ and $\Remark{CI10}$. Because of our $uv$ range flagging before imaging, the directly \incode{tclean}-imaged beam sizes are already close. This convolution further precisely matches $\Remark{CO21}$ and $\Remark{CI10}$ image cube beams. 

Next, moment maps are generated with modules in the PHANGS-ALMA pipeline, consistently for CO and \CI{}. This process includes firstly constructing noise and mask cubes from the image cubes, then combining the masks (i.e., where science signal should present), then computing moment 0 -- line integrated intensity, moment 1 -- line peak velocity, and moment 2 -- line width. The mask cubes are the key in producing moment maps. Here two masks are built: a ``broad mask'' which contains as much plausible science signal as possible, and a ``strict mask'' which is optimized for high-confidence signal. The strict mask is made by finding peaks that have $>4\sigma$ over two consecutive channels then expanding down to $2\sigma$ also with the watershed algorithm. Then the broad mask is obtained by aggregating strict masks made from image cubes convolved to coarser resolutions ($\sim30''$ in this work as well as in \citealt{Leroy2021pipeline}). 
To make consistent moment maps for CO and \CI{}, we aggregate their masks together as a joint mask, for broad and strict masks, respectively, then their final moment maps are computed within the joint mask. 

The resulting $\Remark{CO21}$ and $\Remark{CI10}$ line intensity maps are then used for our next line ratio analysis. The line velocity and line width maps are not further analyzed, but are consistent with the products from the PHANGS-ALMA survey's data release as well as in \citet{Lang2020}. Given that the PHANGS-ALMA 12m+7m+tp $\Remark{CO21}$ has the full spatial information and are deeper, their line velocity and line width maps are more useful for kinematic studies than ours produced from this work.

\section{Assessing the missing flux in the NGC~3627 \CI{} data}
\label{appendix: missing flux}

We assess the level of missing flux in the NGC~3627 \CI{} data in three ways: 
(a) comparing with the \textit{Herschel} SPIRE FTS $\Remark{CI10}$ observation in this galaxy, 
(b) comparing the $uv$-trimmed CO(2-1) moment-0 map with the PHANGS-ALMA short-spacing corrected (12m+7m+tp) moment-0 map, 
and (c) performing CASA simulation and comparing with the ``true'' moment-0 from the skymodel and the derived product from the simulated visibilities. 

Method (a): 
NGC~3627 has been observed with the \textit{Herschel} SPIRE (\citealt{Griffin2010}) Fourier Transform Spectrometer (FTS; \citealt{Naylor2010}) in mapping mode with nearly Nyquist sampling. 
\cite{Liudz2015} reduced the archival data and extracted the line fluxes for each groups of pointings. $\Remark{CI10}$ is detected with $\SNR\sim4-5$ at three sightlines from the galaxy centre to the southern bar end, with R.A. and Dec. coordinates (J2000) 170.061905, +12.991336, 170.066361 +12.987091, and 170.071243 +12.981615, respectively. The $\Remark{CI10}$ luminosities are $1.44 \pm 0.31$, $2.01 \pm 0.48$, and $1.71 \pm 0.40 \times 10^{7} \; \mathrm{K\,km\,s^{-1}\,pc^{2}}$, respectively, where the errors are purely from the RMS of line-free FTS spectra. By convolving our $\Remark{CI10}$ line cube to the FTS beam of $\sim39''$ then extracting the moment maps, we measure a $\Remark{CI10}$ line luminosity of $0.52$, $0.33$ and $0.29 \times 10^{7} \; \mathrm{K\,km\,s^{-1}\,pc^{2}}$ at the above three positions, respectively, which means a missing flux of $36-75\%$ at the galaxy centre and $68-88\%$ at the southern bar end (at a 95\% confidence level). 

Method (b): 
We made ratio maps of our $uv$-trimmed CO(2-1) moment-0 map, and that of the original PHANGS-ALMA short-spacing-corrected (12m+7m+tp) CO(2-1) data but convolved to the corresponding resolution before making the moment maps. The ratio maps are made at two resolutions: the FTS $39''$ beam, and our working resolution of $3.46''$. The former leads to a missing flux of 26\%, 31\% and 36\% from the galaxy centre to the southern bar end. Note that the above \textit{Herschel} FTS analysis indicates in general a higher missing flux, but the FTS measurements are of low $\SNR$ and errors purely from the RMS of the spectra can not fully account for other errors like pointing, spectral response, off-axis spaxel calibration, etc. 
The latter ratio map shows that the missing flux at our working resolution and within the strict mask varies from $\sim$60\% to as low as 6\% at the brightest pixels. This is consistent with our next method based on simulation. 

Method (c): 
We performed detailed simulations of 7m-only $\Remark{CI10}$ and 12m+7m $\Remark{CO21}$ observations with the \incode{CASA} \incode{simobserve} task. We use the PHANGS-ALMA 12m+7m CO(2-1) clean model cube (one of the output products of \incode{CASA} \incode{tclean} runs) as the input \incode{skymodel} for $\Remark{CO21}$, and that scaled by a factor of 0.2 in Kelvin units for $\Remark{CI10}$ (representing a constant $\RCICO=0.2$). 
Note that we do not use the short-spacing-corrected PHANGS-ALMA CO(2-1) cube because it has a finite beam size of $\sim1''$, and simulation based on that may result in more extended emissions. On the contrary, the clean model cube has an infinite resolution, and because both ``multiscale'' and ``singlescale'' cleaning are used in the PHANGS-ALMA imaging, the clean model contains point as well as extended sources. 
We also set the actual observing dates and integration times as the real observations. 
Then we run \incode{CASA} \incode{simobserve} to simulate the visibilities of each observing block, concatenate them together, and process them into image cubes and moment maps in the same way as described in Appendix~\ref{appendix: data reduction}. Meanwhile, by convolving the input \incode{skymodel} with a 2D Gaussian beam corresponding to our working resolution, and integrating over the spectral axis, we obtain a ``true'' \incode{skymodel} moment-0 map which can be directly compared to the imaged moment-0 maps from the simulation. Their ratio map shows a missing flux of $\sim50\%$ to $\sim7\%$ in the strict mask, and more specifically, 24\%, 58\%, 85\% and 96\% pixels selected in our method as for real data have a missing flux $< 20\%$, $< 30\%$, $< 40\%$ and $< 50\%$, respectively. 

\begin{figure*}
\centering%
\includegraphics[width=\textwidth, trim=0 7mm 0 0, clip]{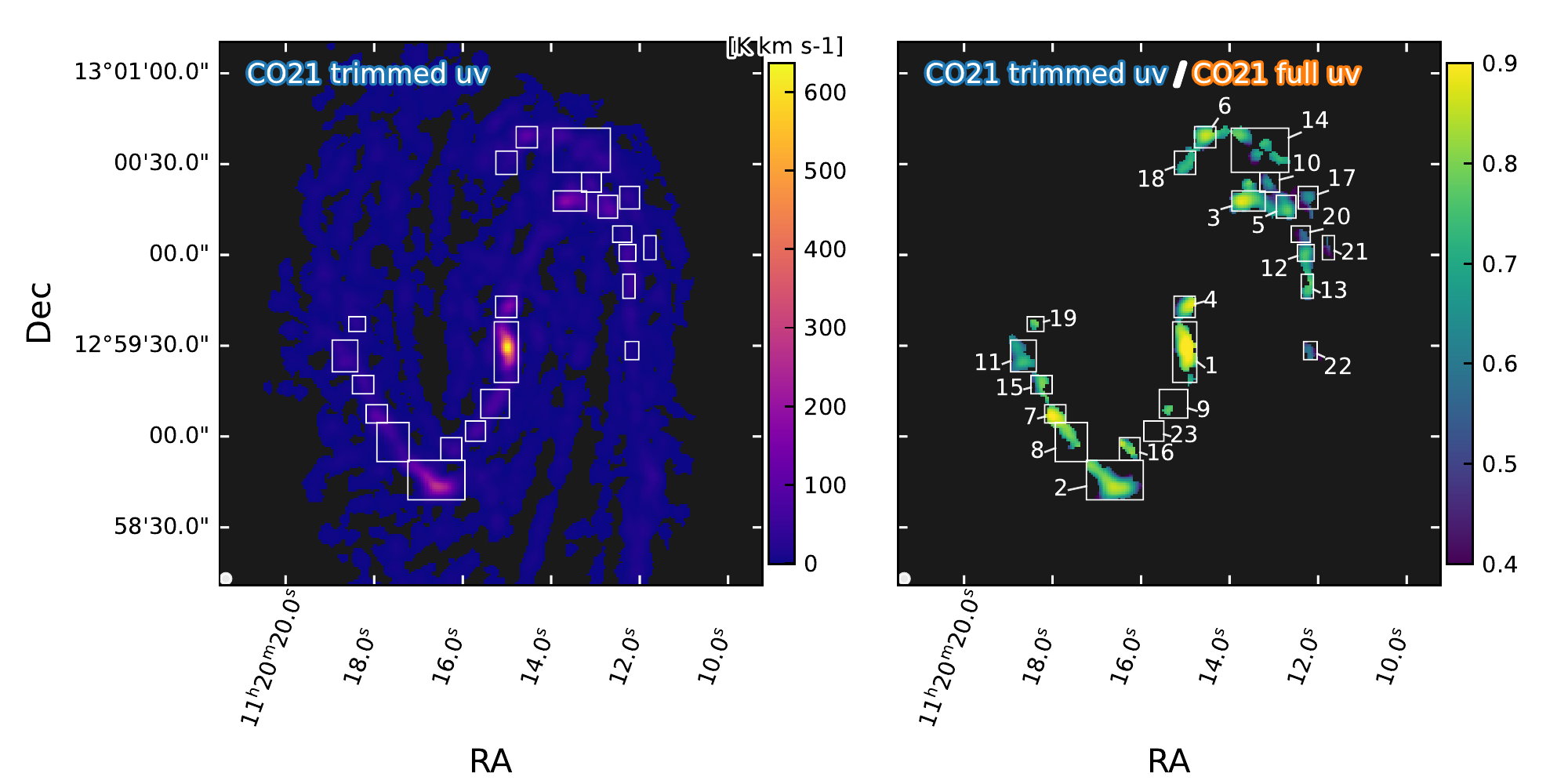}
\includegraphics[width=\textwidth, trim=0 0 0 5mm, clip]{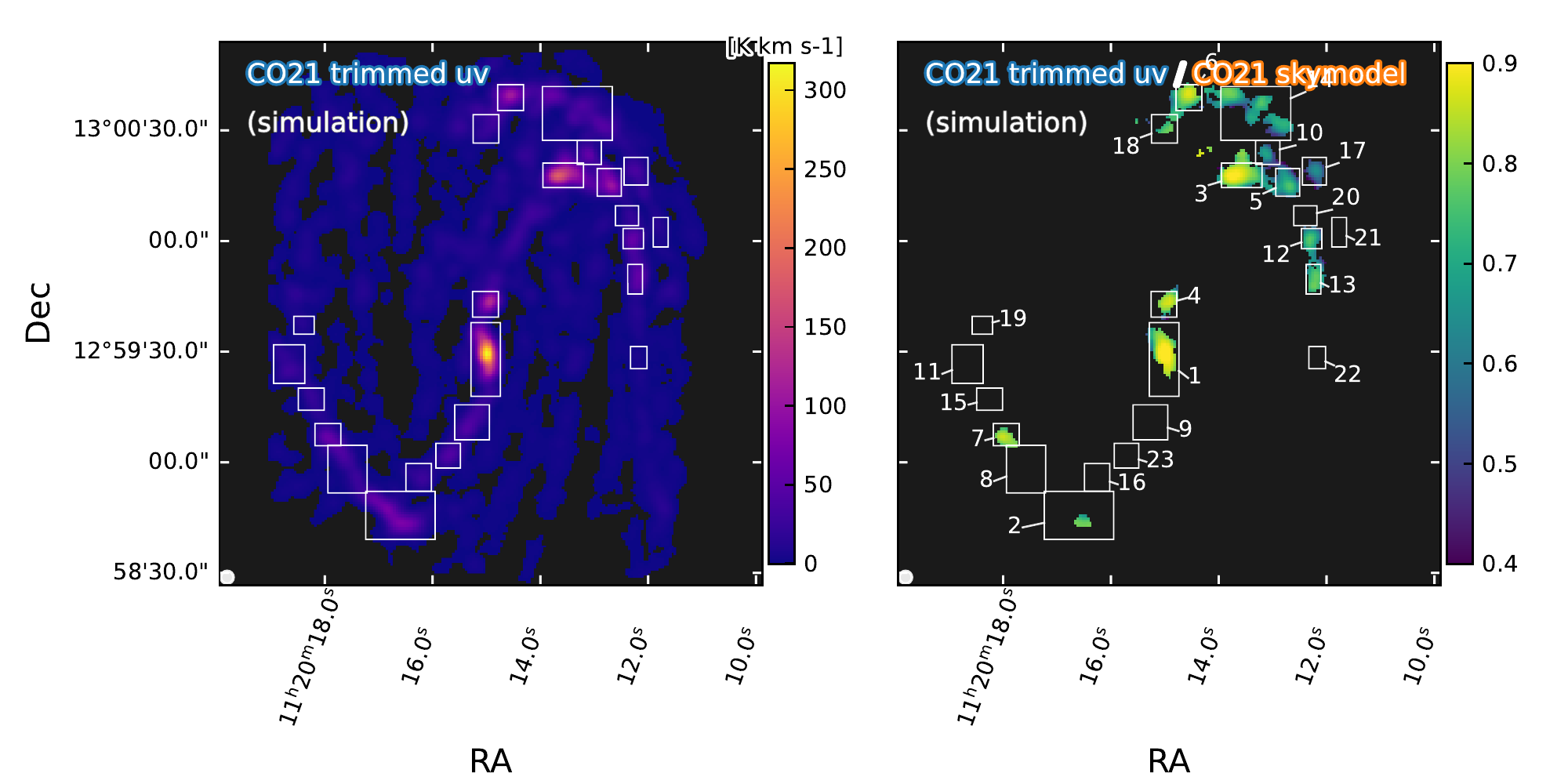}
\caption{%
Upper left panel: short-spacing corrected (12m+7m+tp) PHANGS-ALMA CO(2-1) derived moment-0 map. 
Upper right panel: ratio map of the $uv$-trimmed CO(2-1) moment-0 map used in this work to that in the left panel. 
Lower left panel: the CO(2-1) moment-0 map derived from our simulation with $uv$ trimming as for real data. 
Lower right panel: ratio map of the $uv$-trimmed CO(2-1) moment-0 map based on the simulation to the ``true'' \incode{skymodel} moment-0 map as shown left. 
All moment maps are derived by first convolving cubes to our working resolution of $3.46''$ then performing masking analysis as described in Appendix~\ref{appendix: data reduction}, except for the ``true'' \incode{skymodel} moment-0 map which is computed without any masking. 
Note that in the simulation based analysis, some regions have too low $\SNR$ due to their faint CO brightness or insufficient cleaning, thus they are out of the mask and show blank line ratios (instead of very low missing flux) in the lower right panel. 
\label{fig: missing flux}
}
\end{figure*}

Fig.~\ref{fig: missing flux} illustrates the results from methods (b) (upper panels) and (c) (lower panels). Left panels show the $\Remark{CO21}$ moment-0 maps, analyzed at the same working resolution as our main analysis in Appendix~\ref{appendix: data reduction}. The simulated data show a fainter emission because we used the clean mode cube as the \incode{skymodel}, which was made by cleaning the PHANGS-ALMA 12m+7m data down to 1~$\sigma$ within the clean mask voxels. This will indeed lose some fluxes. However, the right panels show that the two methods lead to consistent missing flux results. 

Note that in the last panel of Fig.~\ref{fig: missing flux}, some regions appear black in the simulation-based analysis because we require line ratios to be analyzed in the strict mask, while these regions are not in it anymore. This is because the \incode{skymodel} loses some fluxes as mentioned above, thus the strict mask is smaller, and also because in the simulation we assumed a constant $\RCICO$ which may lead to fainter \CI{} than real data along faint sightlines, e.g., Region 22,  thus further shrinking the strict mask where both lines need to have sufficient $\SNR$. 

Bearing these in mind, we conclude that methods (b) and (c) have consistent results on the missing flux, and method (a) is less accurate due to the unconsidered uncertainties. 

These checks show that the overall missing flux in NGC~3627 is 31\% based on the full- and trimmed-$uv$ $\Remark{CO21}$ analysis, or $\sim36-88\%$ based on the more-uncertain \textit{Herschel} data. At the pixel level, it varies from 7\% to $\sim$50\% from bright to faint sightlines at our working resolution of $3.46''$. 
These values are as expected given our ACA-only observation setup. 
We emphasize again that by applying the $uv$-matching in our data reduction, our \CI{}/\CO{} line ratio study is not obviously affected by the missing flux issue, except perhaps in the faintest pixels or ``cloud edges''.

\section{Assessing noise-induced \CI{} enhancement with simulation}
\label{appendix: simulation}

In Appendix~\ref{appendix: missing flux} we described our simulation. We also use it to assess how the noise can induce artificial [\CI/CO] line ratio variation. 

\begin{figure*}
\centering%
\includegraphics[width=\textwidth]{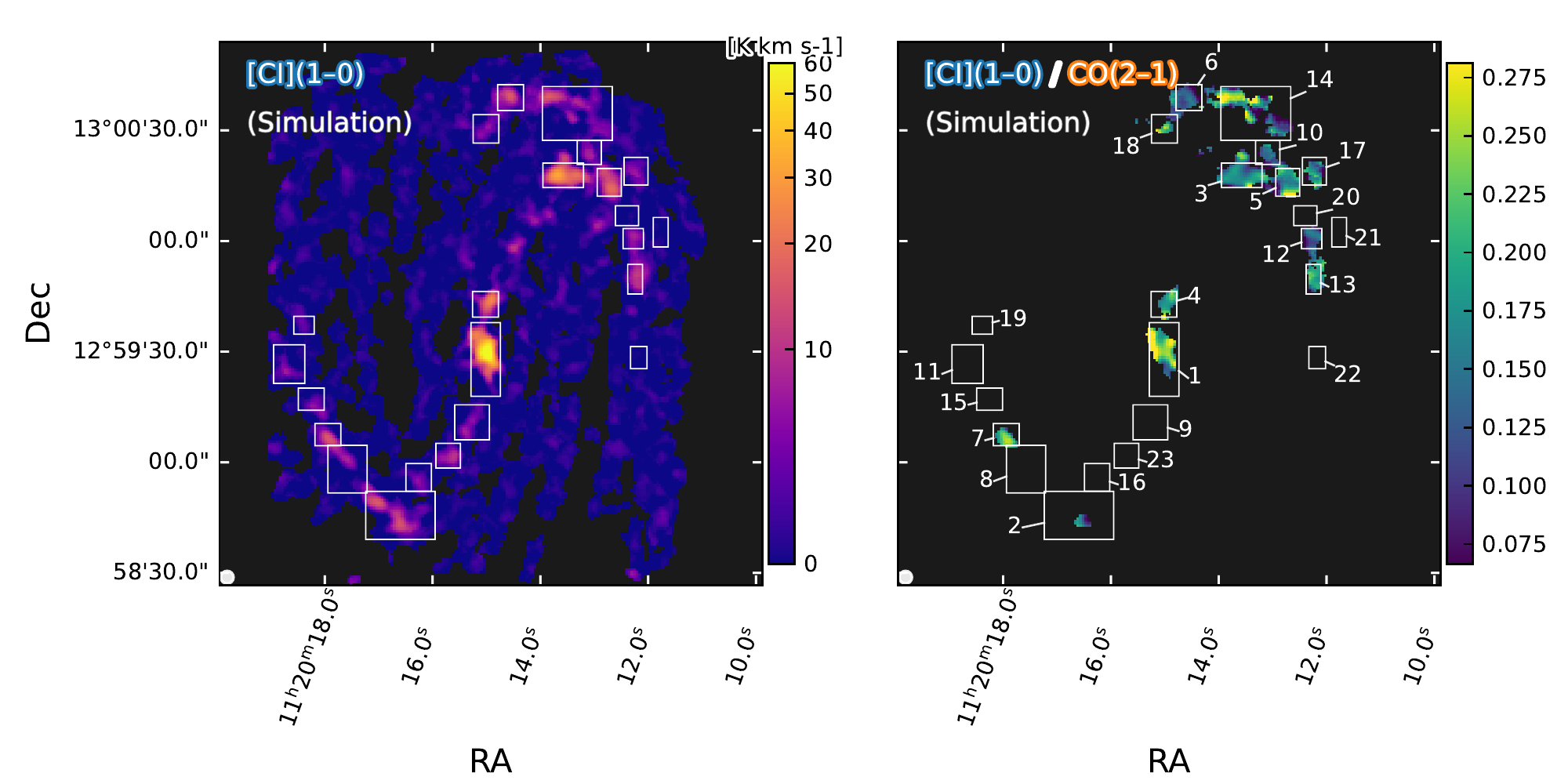}
\caption{%
\CI{} line flux and $\RCICO$ line ratio image from the simulation mimicking real observation but with a constant input line ratio $\RCICO=0.2$. 
}
\label{fig: sim line ratio image}
\end{figure*}

\begin{figure*}
\centering%
\includegraphics[width=0.9\textwidth]{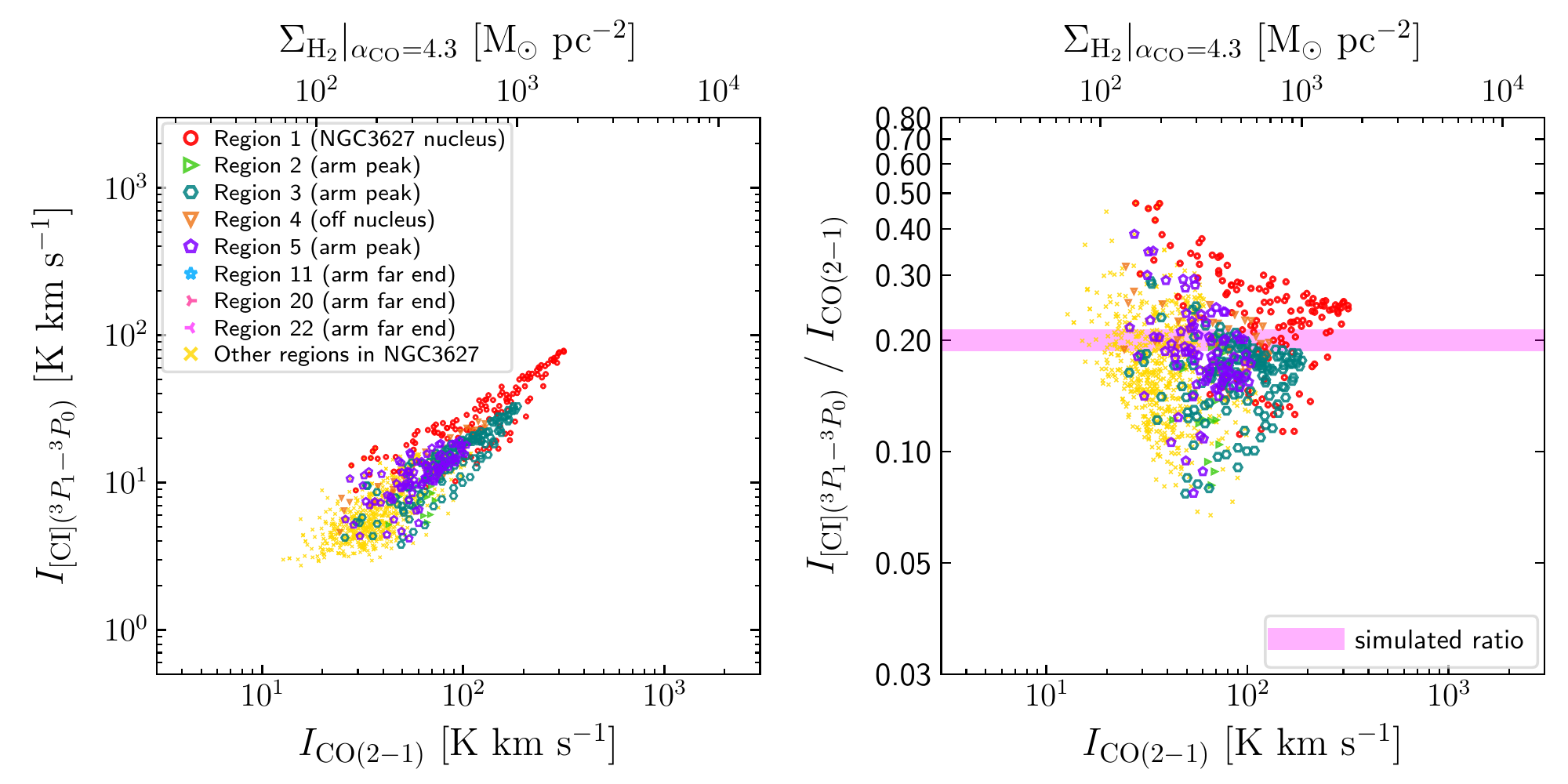}
\includegraphics[width=0.9\textwidth, trim=0 0 0 13mm, clip]{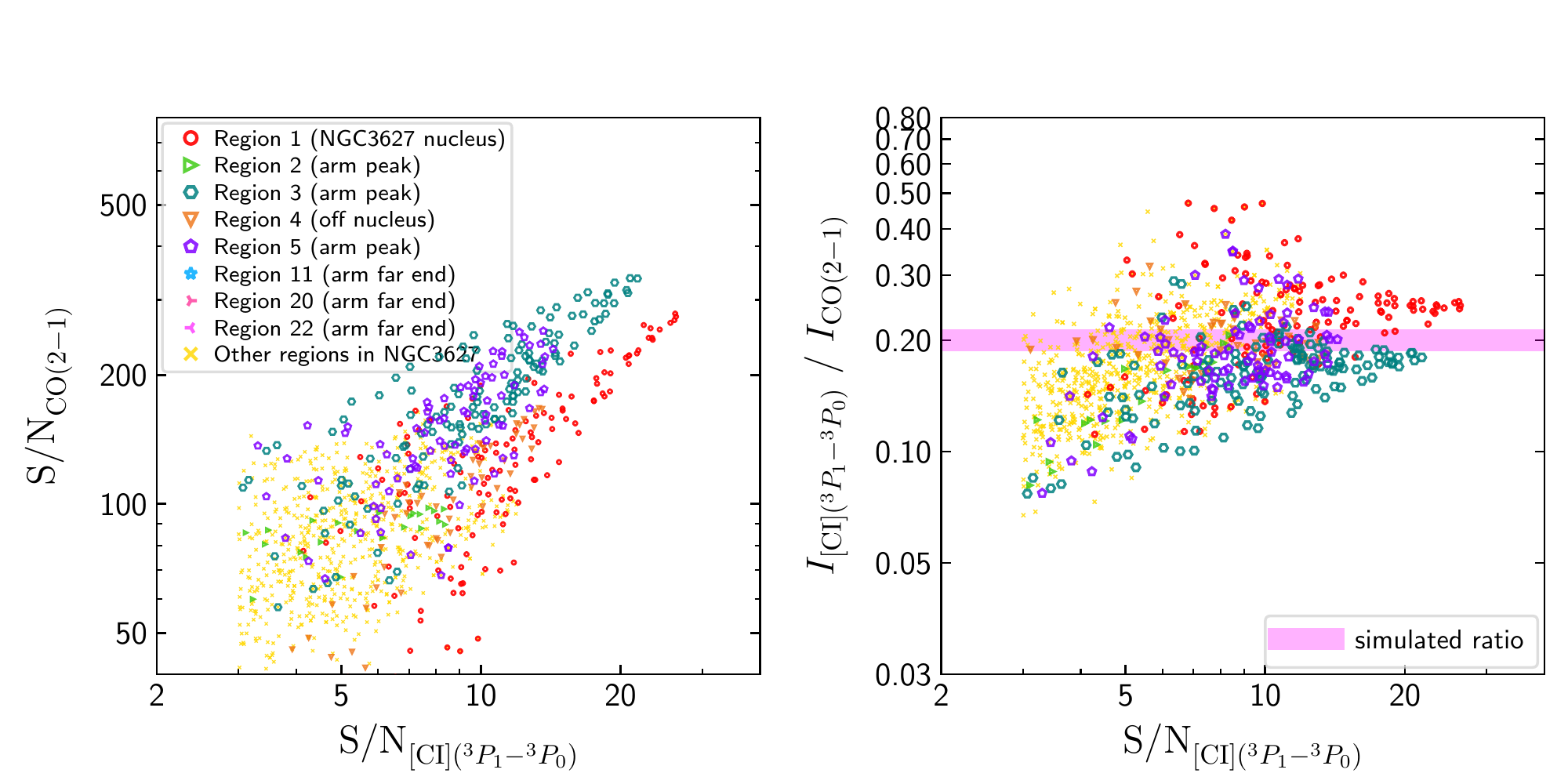}
\caption{%
\textit{Upper panels}: $\Remark{CI10}$ and $\Remark{CO21}$ line brightness and ratio similar to Fig.\ref{fig: line ratio vs CO} but from our simulation (with a constant input $\RCICO=0.2$ as indicated by the horizontal shading in the right panels). 
\textit{Lower panels}: Similar to the upper panels but showing line $\SNR$ instead of brightness. This illustrates how $\RCICO$ can be biased in low-$\SNR$ situation. 
}
\label{fig: sim line ratio scatter}
\end{figure*}

In Fig.~\ref{fig: sim line ratio image} we show the line ratio map from the simulation with a constant $\RCICO=0.2$, and in Fig.~\ref{fig: sim line ratio scatter} we show the scatter plots with simulation results similar to Fig.~\ref{fig: line ratio vs CO}. Interestingly, we see certain \CI/CO variations, especially \CI{} ``enhancement'' at the edge of galaxy centre's gas clump (region~1), and the edge of the northern spiral arm (Region 14). Nevertheless, the trend is very different from the real data, i.e., Fig.~\ref{fig: line ratio vs CO}. 
First, galaxy centre has a much weaker [\CI/CO] in the real data, whereas in the simulation it is inverted. 
Second, the scatter is much larger in real data, with a tail of \CI{} enhanced regions which is not seen in the simulation. 
Third, from the bottom panels of Fig.~\ref{fig: sim line ratio scatter}, where $\SNR$ is shown, we see that $\RCICO$ is biased to be lower in low-$\SNR$ situation. Thus the high \CI{} in real data is unlikely to be a bias of low $\SNR$. 

\section{\CI\ and \CO\ spectra of selected regions in NGC~3627}
\label{appendix: all regions}

\begin{figure*}
\centering%
\includegraphics[width=0.7\textwidth]{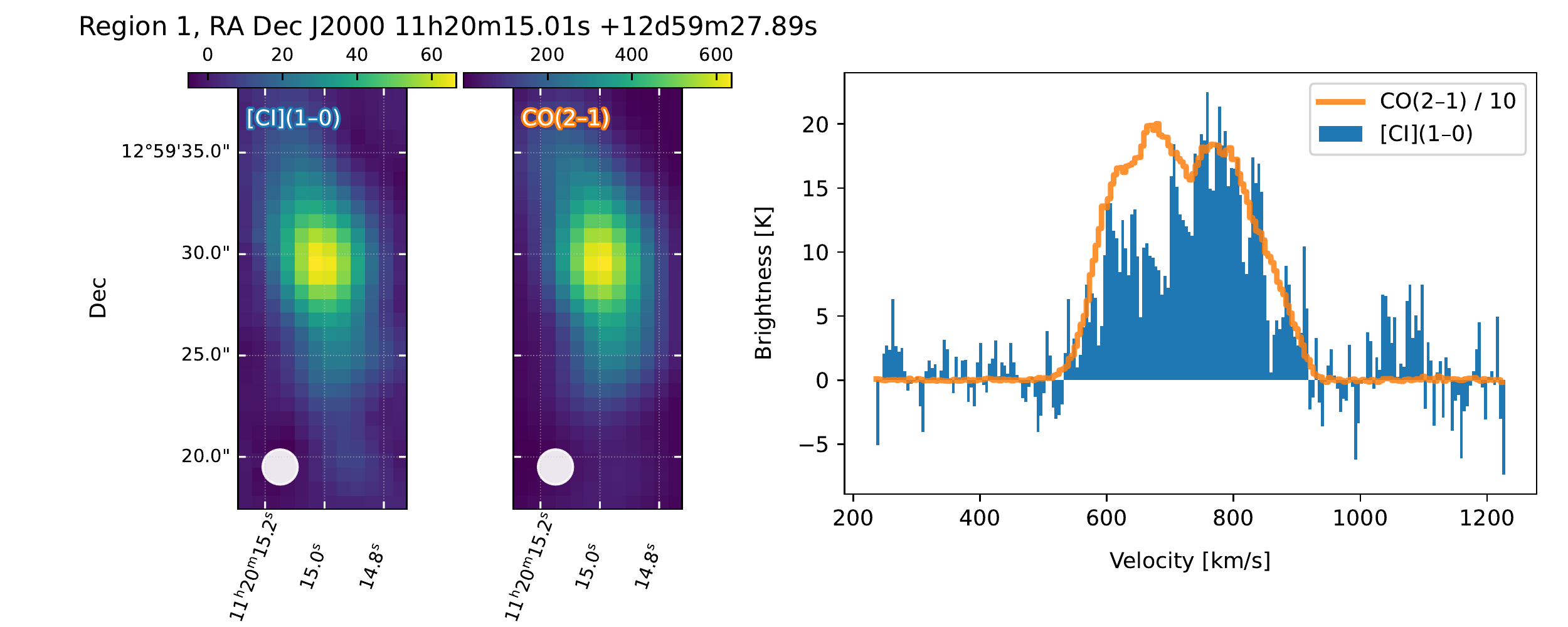}
\includegraphics[width=0.7\textwidth]{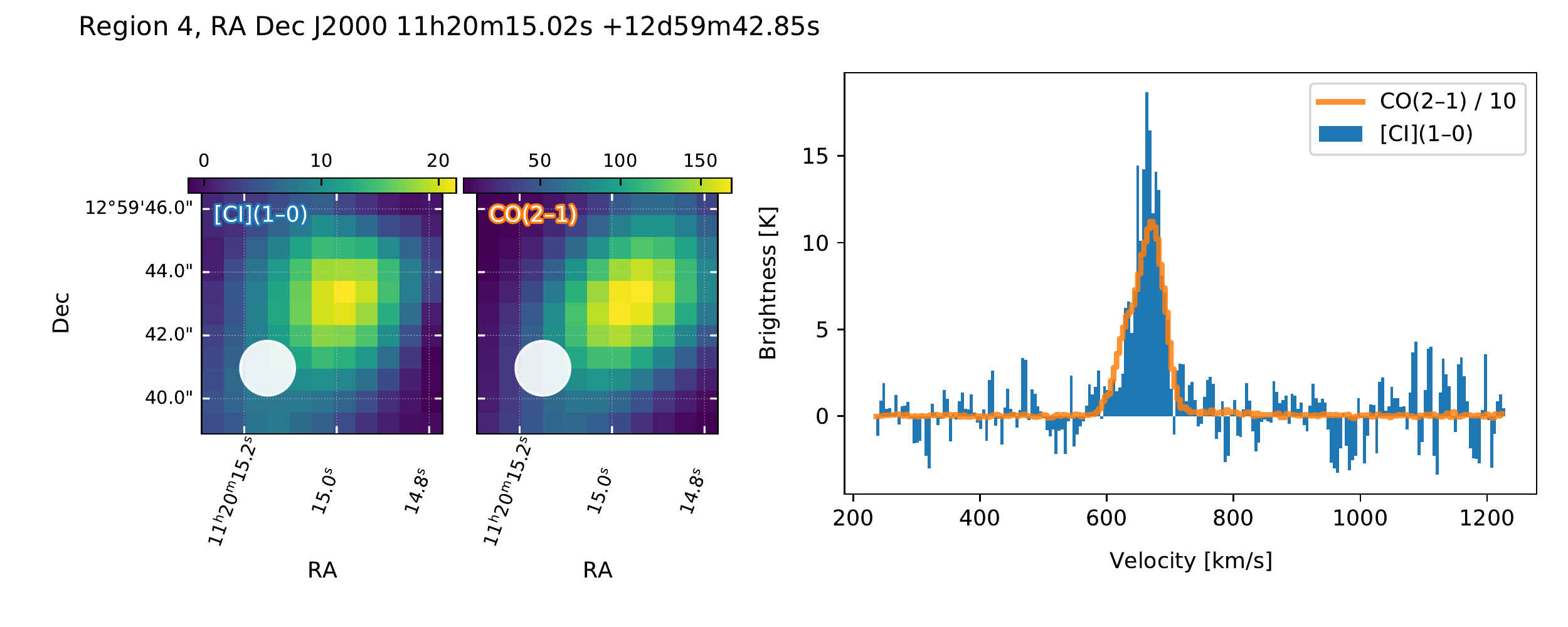}
\includegraphics[width=0.7\textwidth]{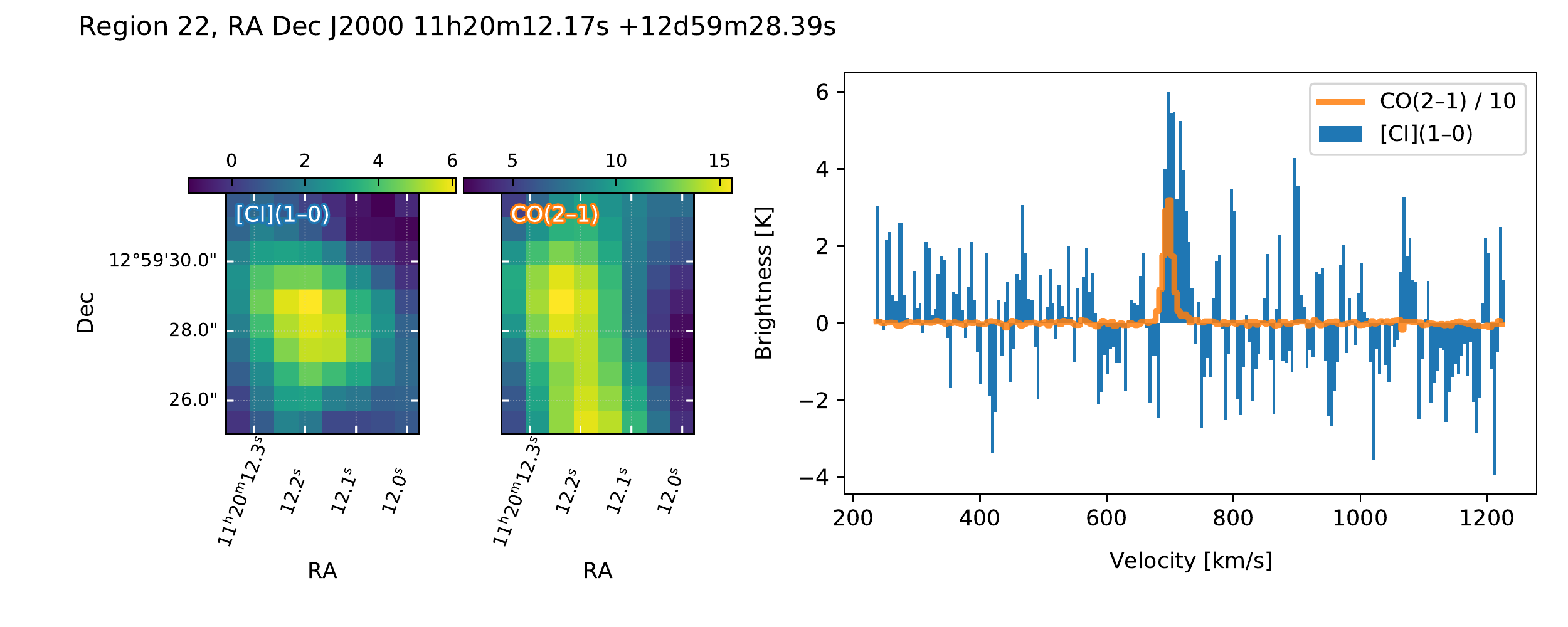}
\caption{$\Remark{CI10}$ and $\Remark{CO21}$ moment-0 maps and spectra of Regions 1, 4 and 22 as defined in Fig.~\ref{fig: NGC3627 line ratio map}. All regions' spectra are available in our online data release (see Sect.~\ref{subsection: ALMA Band 8 Observation}).  
\label{fig: NGC3627 regions}
}
\end{figure*}

We show the examples of $\Remark{CI10}$ and $\Remark{CO21}$ moment-0 maps and spectra for Regions 1, 4 \& 22 in Fig.~\ref{fig: NGC3627 regions}. They represent the NGC~3627 centre, off-centre location and the location at the outer disk of the gaseous spiral arm as discussed in Sect.~\ref{sec: Results}. 

\end{appendix}

\end{document}